\documentclass[twocolumn,showpacs,preprintnumbers,superscriptaddress,amsmath,amssymb]{revtex4}
\usepackage{amsmath, amsthm}
\usepackage{graphicx}
\usepackage{dcolumn}
\usepackage{bm}
\usepackage{amssymb}
\usepackage{natbib}

\makeatletter
\@ifundefined{textcolor}{}
{%
 \definecolor{BLACK}{gray}{0}
 \definecolor{WHITE}{gray}{1}
 \definecolor{RED}{rgb}{1,0,0}
 \definecolor{GREEN}{rgb}{0,1,0}
 \definecolor{BLUE}{rgb}{0,0,1}
 \definecolor{CYAN}{cmyk}{1,0,0,0}
 \definecolor{MAGENTA}{cmyk}{0,1,0,0}
 \definecolor{YELLOW}{cmyk}{0,0,1,0}
 }

\newcommand{\HF}{HF}
\newcommand{\HFB}{HFB}
\newcommand{\Tr}{{\rm Tr}}
\newcommand{\cH}{{\cal H}}
\newcommand{\cR}{{\cal R}}
\newcommand{\cU}{{\cal U}}
\newcommand{\cT}{{\cal T}}
\newcommand{\cG}{{\cal G}}
\newcommand{\cQ}{{\cal Q}}
\newcommand{\cN}{{\cal N}}

\newcommand{\thalf}{\tfrac{1}{2}}

\begin{document}

\title{Polarization corrections to single-particle energies studied within the
energy-density-functional and QRPA approaches}

\author{D. Tarpanov}
\affiliation{Institute of Theoretical Physics, Faculty of Physics, University of Warsaw,
ul. Ho\.za 69, PL-00-681 Warsaw, Poland}
\affiliation{Institute for Nuclear Research and Nuclear Energy, 1784 Sofia,
Bulgaria}

\author{J. Toivanen}
\affiliation{Department of Physics, P.O. Box 35 (YFL),
University of Jyv\"askyl\"a, FI-40014  Jyv\"askyl\"a, Finland}

\author{J. Dobaczewski}
\affiliation{Institute of Theoretical Physics, Faculty of Physics, University of Warsaw,
ul. Ho\.za 69, PL-00-681 Warsaw, Poland}
\affiliation{Department of Physics, P.O. Box 35 (YFL),
University of Jyv\"askyl\"a, FI-40014  Jyv\"askyl\"a, Finland}

\author{B.G. Carlsson}
\affiliation{Division of Mathematical Physics, LTH, Lund University, Post Office Box 118,
S-22100 Lund, Sweden}

\begin{abstract}
\begin{description}
\item[Background:]
Models based on using perturbative polarization corrections
and mean-field blocking approximation give conflicting results for
masses of odd nuclei.

\item[Purpose:] Systematically investigate the polarization and mean-field
models, implemented within self-consistent approaches that use
identical interactions and model spaces, so as to find reasons for
the conflicts between them.

\item[Methods:] For density-dependent interactions and with pairing
correlations included, we derive and study links between the
mean-field and polarization results obtained for energies of odd
nuclei. We also identify and discuss differences between the
polarization-correction and full particle-vibration-coupling (PVC) models.
Numerical calculations are performed for the mean-field ground-state
properties of deformed odd nuclei and then compared to the polarization
corrections determined by using the approach that conserves spherical
symmetry.

\item[Results:]
We have identified and numerically evaluated self-interaction (SI) energies
that are at the origin of different results obtained within the
mean-field and polarization-correction approaches.

\item[Conclusions:]
Mean-field energies of odd nuclei are polluted by the SI energies, and
this makes them different from those obtained by using
polarization-correction methods. A comparison of both approaches
allows for the identification and determination of the SI terms,
which then can be calculated and removed from the mean-field results,
giving the self-interaction-free energies. The simplest deformed
mean-field approach that does not break parity symmetry is unable to
reproduce full PVC effects.

\end{description}
\end{abstract}
\pacs{21.10.Pc,21.60.Jz}

\maketitle

\section{Introduction}

The perturbative particle-vibration-coupling (PVC) model for
odd-particle-number nuclei emerges naturally from the self-consistent
Green's function theory~\cite{[Dic05]}. It describes the polarization
of the nucleus when one particle is added or removed~\cite{[Boh75w]},
and its results can, in principle, be directly compared against
experimental data. As used in nuclear physics, the perturbative PVC
method employs one-particle or one-hole states (or one-quasiparticle
states) coupled with the RPA or QRPA excitations of an even-even
reference nucleus, and the residual nucleon-nucleon interaction that mixes
these states at second-order perturbation theory. Numerous PVC
calculations of increasing level of sophistication have already been
performed, see, e.g.,
Refs.~\cite{(Ham74),[Ber80w],[Mah85],(Gal88d),[Slu93],[Avd99a],[Bar06],(Lit07),%
(Mis08),(Yos09),(Col01),(Col10),(Lit11),(Lit12),(Miz12),[Ind12],(Som13)}
and excellent recent reviews thereof available in
Refs.~\cite{(Col10),[Ind12],[Bor10]}.

An alternative to describing odd nuclei within the perturbative PVC
calculations are the energy-density-functional (EDF) methods, see,
e.g.,
Refs.~\cite{[Rut98],[Sat98a],[Dug02a],[Dug02b],[Zal08],[Ber09a],[Sch10]},
which use blocking of single-particle (s.p.) or quasiparticle
orbitals. To distinguish these methods from the full PVC approach, in
the present study we call them mean-field polarizations or
polarization corrections. The advantage of blocked mean-field
calculations is that they are non-perturbative and variational.

As it turns out, the effects obtained within the blocked mean-field
methods are substantially different and, in general, weaker than
those obtained from the perturbative PVC~\cite{[Bor10]}. This
discrepancy between models, even when using exactly the same
interactions and model spaces, needs to be solved, and this is the
main purpose of the present work.

The link between the mean-field and perturbative methods has been
proposed long time ago~\cite{[Bro70],[Bla86],[Lip87],[Lip89]}. Here,
we identify several approximations that are required to firmly
establish such a link, and we also extend the derivations to EDFs
based on density-dependent interactions and to those that include
pairing correlations. Since the determination of mean-field
polarizations requires breaking symmetries, no numerical comparison
of the two approaches, such as given here, is available up to now.
As required by a thorough comparison, both in the mean-field and (Q)RPA
calculations we use full self-consistency and exactly the same
particle-hole EDFs, pairing interactions, and model spaces.

The paper is built around two main chapters presenting theoretical
derivations in Sec.~\ref{sec2} and Appendix~\ref{sec5}, and numerical
results in Sec.~\ref{sec3}. For theory, we present results pertaining
to the Hartree-Fock (HF) approximation (Sec.~\ref{sec2.2}), density
functionals (Sec.~\ref{sec2.4}), and pairing correlations treated
within the Hartree-Fock-Bogoliubov (\HFB) framework
(Sec.~\ref{sec2.3}). Then, results of calculations are discussed for
the density-independent (Sec.~\ref{sec3.1}) and density-dependent
(Sec.~\ref{sec3.2}) particle-hole interactions, and for paired
systems (Sec.~\ref{sec3.3}). Conclusions are given in
Sec.~\ref{sec4}.

\section{Theory}
\label{sec2}

\subsection{Polarization corrections in the {\HF} approximation}
\label{sec2.2}

In this section we revisit the classic
problem~\cite{[Bro70],[Bla86],[Lip87],[Lip89]} of the polarization
effect exerted by an odd particle on a mean-field state. To put
further discussion in perspective, we study the problem in the {\HF}
approximation, and we assume that the mean field is obtained by
the {\HF} averaging~\cite{[RS80]} of a given known two-body
{\em density-independent} interaction that has antisymmetrized matrix elements
${\bar{v}}^{}_{i'k'ik}$.

Let $\rho^A$ and $h^A$ denote, respectively, the self-consistent density matrix
and {\HF} Hamiltonian for a system of $A$ fermions: $\Tr\rho^A$=A$, \left\lbrack
h^A, \rho^A \right\rbrack=0$. Similarly, let $\rho^{A\pm1}$
denote the self-consistent density-matrices corresponding to the
{\HF} solution for the $(A\pm1)$-particle system:  $\Tr\rho^{A\pm1}=A\pm1$,
$\left\lbrack{}h^{A\pm1}, \rho^{A\pm1} \right\rbrack=0$. We
use the notation, in which the upper and lower signs correspond to
adding or subtracting a particle. Without pairing correlations, even and
odd systems are described in exactly the same way, so without any
loss of generality, we assume that $A$ is even.

Self-consistent {\HF} total energies of the $A$- and $(A\pm1)$-particle systems
are given by~\cite{[RS80]},
\begin{subequations}
  \label{eq:50}
\begin{eqnarray}
  \label{eq:50a}
  E^{A} &=& {\rm Tr}(t\rho^{A}) +
  \thalf {\Tr}_1{\Tr}_2(\rho^{A}\bar{v}\rho^{A}) , \\
  \label{eq:50b}
  E^{A\pm1} &=& {\rm Tr}(t\rho^{A\pm1}) +
  \thalf {\Tr}_1{\Tr}_2(\rho^{A\pm1}\bar{v}\rho^{A\pm1}) .
\end{eqnarray}
\end{subequations}
Here, $t$ represents the matrix of one-body kinetic energy. In what
follows {\em we always neglect} the so-called center-of-mass
correction to the kinetic energy~\cite{[RS80],[Ben03]}. These
corrections are explicitly $A$-dependent and thus give trivial
so-called mass polarization corrections~\cite{[Zal08]} to energy
differences $E^{A\pm1}-E^{A}$. Although they can always be added, they
would obscure the analysis of standard polarization corrections,
which are due to two-body interactions, and which are the main focus
of the present study.

Suppose now that $\rho^\lambda$ ($\Tr\rho^\lambda=1$) is the density
matrix of a s.p.\ state $\lambda$. We may now ask
what the relations are between the density matrices
$\rho^A$, $\rho^{A\pm1}$, and $\rho^\lambda$. Of course, we can
always define a corrective density matrix $\delta\rho$
($\Tr\delta\rho=0$) such that, by definition,
  \begin{equation}
    \label{eq:1}
    \rho^{A\pm1} = \rho^A \pm \rho^\lambda + \delta\rho .
  \end{equation}
However, a perturbative treatment can only be obtained in the case
when $\delta\rho$ is small -- small in the sense that when the energy
of the odd systems $E^{A\pm1}$, Eq.~(\ref{eq:50b}), is calculated for the
density matrix in Eq.~(\ref{eq:1}), only terms up to second order in
$\delta\rho$ are important.

Note that, by definition, the three
density matrices are hermitian and projective,
\begin{subequations}
\begin{eqnarray}
  \label{eq:49a}
\left(\rho^{A}    \right)^2&=&\rho^{A}    =\left(\rho^{A}    \right)^+, \\
 \label{eq:49b}
\left(\rho^{A\pm1}\right)^2&=&\rho^{A\pm1}=\left(\rho^{A\pm1}\right)^+, \\
 \label{eq:49c}
\left(\rho^\lambda\right)^2&=&\rho^\lambda=\left(\rho^\lambda\right)^+.
\end{eqnarray}
\end{subequations}
Also note that $\delta\rho$ does depend on the polarizing orbital
$\lambda$; nevertheless, we do not mark it with superscript
$\lambda$. This is to avoid a confusion of understanding $\delta\rho$
as a correction to the orbital itself; indeed, this correction
certainly corresponds to a modification of {\em all} orbitals of the
system.

\subsubsection{Properties of $\delta\rho$}
\label{sec2.2.1}
The corrective density matrix $\delta\rho$ can be small only when the
orbital $\lambda$, and the states in even and odd nuclei, are chosen in a
specific way. We may then have four interesting cases to consider.
In the first case, let us assume that we initially solve the self-consistent
equations of the even $A$-particle system, and $|\lambda\rangle$ is one of
the unoccupied {\HF} eigenstates therein (a particle state), that is,
$h^A|\lambda\rangle=e_\lambda|\lambda\rangle$,
$\rho^{A}\rho^\lambda=0$. We may now put a particle in this orbital
and solve the self-consistent equations of the $(A+1)$-particle
system. In this sense, the $(A+1)$-particle system becomes polarized
by an addition of a particle to the $A$-particle system. Note that by
this procedure {\em all} {\HF} single-states of the $A$-particle
system become modified, including the added orbital
$|\lambda\rangle$.

The second case is obtained by a similar procedure, where instead we arrive at a polarized
$(A-1)$-particle system. For that, we pick $|\lambda\rangle$ as one of
the occupied {\HF} eigenstates of the $A$-particle system (a hole
state), that is, $h^A|\lambda\rangle=e_\lambda|\lambda\rangle$,
$\rho^{A}\rho^\lambda=\rho^\lambda$. By removing a particle from this
state and solving the self-consistent equations of the $(A-1)$-particle
system, we now obtain the polarization correction corresponding to a
hole. Note that for the two choices discussed up to now, the density
matrices $\rho^{A}+\rho^\lambda$ and $\rho^{A}-\rho^\lambda$,
are projective, that is,
\begin{equation}
  \label{eq:50c}
(\rho^{A}\pm\rho^\lambda)^2 = \rho^{A}\pm\rho^\lambda .
\end{equation}

The two remaining interesting cases correspond to inverse
polarizations, namely, we may initially solve the self-consistent
equations of the $(A\pm1)$-particle systems, and then pick $\lambda$
either as an unoccupied orbital in the $(A-1)$-particle system,
$h^{A-1}|\lambda\rangle=e_\lambda|\lambda\rangle$,
$\rho^{A-1}\rho^\lambda=0$ or as an occupied orbital in the
$(A+1)$-particle system,
$h^{A+1}|\lambda\rangle=e_\lambda|\lambda\rangle$,
$\rho^{A+1}\rho^\lambda=\rho^\lambda$. Of course, in both cases, the
self-consistent equations solved for the $A$-particle system give the
same solutions as before, however, now orbitals $|\lambda\rangle$
correspond to the $(A\pm1)$-particle systems, and thus density
matrices $\rho^{A-1}+\rho^\lambda$ and $\rho^{A+1}-\rho^\lambda$, are
projective, that is,
\begin{equation}
  \label{eq:51}
(\rho^{A\mp1}\pm\rho^\lambda)^2 = \rho^{A\mp1}\pm\rho^\lambda .
\end{equation}

We see that equations we are going to derive for the corrective
density $\delta\rho$ do depend on the choices made for the
orbital $|\lambda\rangle$. For the direct polarizations, that is,
when $|\lambda\rangle$ is a self-consistent orbital in the even system,
we square both sides of Eq.~(\ref{eq:1}), and from Eq.~(\ref{eq:50c})
we obtain
\begin{eqnarray}
  \label{eq:j4}
      \delta\rho &=&
                   (\rho^A\pm\rho^\lambda)\delta\rho +
                   \delta\rho(\rho^A\pm\rho^\lambda) + (\delta\rho)^2  .
\end{eqnarray}
For the inverse polarizations, that is,
when $|\lambda\rangle$ are self-consistent orbitals in the odd systems,
we rewrite Eq.~(\ref{eq:1}) in the form,
  \begin{equation}
    \label{eq:1a}
    \rho^{A\pm1} \mp \rho^\lambda = \rho^A  + \delta\rho ,
  \end{equation}
and then square both sides, which from Eq.~(\ref{eq:51}) gives,
\begin{eqnarray}
  \label{eq:j4a}
      \delta\rho &=&
                   \rho^A\delta\rho +
                   \delta\rho\rho^A + (\delta\rho)^2  .
\end{eqnarray}

Equations (\ref{eq:j4}) and (\ref{eq:j4a}) allow us to derive
specific properties of $\delta\rho$ that, however, are different
for direct and inverse polarizations.
Assuming that we can split $\delta\rho$ into terms
of first, second, and higher (neglected) orders, that is,
\begin{eqnarray}
  \label{eq:52}
      \delta\rho &=& \delta\rho^{(1)} + \delta\rho^{(2)} + \ldots,
\end{eqnarray}
we now separately discuss direct and inverse polarizations.
In what follows, we refer to the
expansion in Eq.~(\ref{eq:52}) as RPA expansion, and we strive to discuss
what an acceptable magnitude of $\rho^\lambda$ is, for which such an
expansion is meaningful.

Beginning with the inverse polarizations, Eq.~(\ref{eq:j4a}) gives
\begin{subequations}
\begin{eqnarray}
  \label{eq:53a}
      \delta\rho^{(1)} &=& \rho^A\delta\rho^{(1)} + \delta\rho^{(1)}\rho^A , \\
  \label{eq:53b}
      \delta\rho^{(2)} &=& \rho^A\delta\rho^{(2)} + \delta\rho^{(2)}\rho^A + (\delta\rho^{(1)})^2.
\end{eqnarray}
\end{subequations}
We can now discuss properties of the particle-hole (ph),
particle-particle (pp), and hole-hole (hh) matrix elements of
$\delta\rho$, where the standard hole and particle states correspond
to the occupied and unoccupied states, respectively, {\em in the even
$A$-particle system}, that is,
\begin{eqnarray}
  \label{eq:53c}
       \rho^A_{hh'} &=& \delta_{hh'}, \quad
       \rho^A_{ph}   = 0,  \quad
       \rho^A_{hp}   = 0,  \quad
       \rho^A_{pp'}  = 0.
\end{eqnarray}
Equation (\ref{eq:53a}) does not put any
constraint on the ph matrix elements of $\delta\rho^{(1)}$, and it requires
that its pp and hh matrix elements vanish identically. Therefore, the
leading-order (second-order) pp and hh matrix elements of
$\delta\rho$ are determined by Eq.~(\ref{eq:53b}), and they solely
depend on the leading-order (first-order) ph matrix elements thereof, that
is,
\begin{subequations}
  \label{eq:57}
\begin{eqnarray}
  \label{eq:57a}
      \delta\rho^{(2)}_{pp'} &=&
                    \sum_h \delta\rho^{(1)}_{ph} \delta\rho^{(1)}_{hp'} , \\
  \label{eq:57b}
      \delta\rho^{(2)}_{hh'} &=&
                  - \sum_p \delta\rho^{(1)}_{hp} \delta\rho^{(1)}_{ph'} .
\end{eqnarray}
\end{subequations}
We see that the standard ph structure of $\delta\rho$, pertaining to
the $A$-particle system, appears for the {\em inverse polarizations}.
However, as we derived above, in this case the polarizing orbital
$|\lambda\rangle$ must be calculated in the {\em odd} system.

Let us next discuss direct polarizations, for which Eq.~(\ref{eq:j4}) holds.
The RPA expansion (\ref{eq:52}) then gives
\begin{subequations}
\begin{eqnarray}
  \label{eq:54a}
      \delta\rho^{(1)} &=& (\rho^A\pm\rho^\lambda)\delta\rho^{(1)}
                           + \delta\rho^{(1)}(\rho^A\pm\rho^\lambda) , \\
  \label{eq:54b}
      \delta\rho^{(2)} &=& (\rho^A\pm\rho^\lambda)\delta\rho^{(2)}
                           + \delta\rho^{(2)}(\rho^A\pm\rho^\lambda)
\nonumber \\ &&
                           + (\delta\rho^{(1)})^2.
\end{eqnarray}
\end{subequations}
Properties of the particle-hole (PH),
particle-particle (PP), and hole-hole (HH) matrix elements of
$\delta\rho$, now pertain to nonstandard hole and particle states, which are
the occupied and unoccupied states, respectively, {\em in the odd
$(A\pm1)$-particle system}, that is,
\begin{eqnarray}
  \label{eq:54c}
       (\rho^A\pm\rho^\lambda)_{HH'} &=& \delta_{HH'}, \quad
       (\rho^A\pm\rho^\lambda)_{PH}   = 0,
\nonumber \\
       (\rho^A\pm\rho^\lambda)_{HP}  &=& 0,  \quad
       (\rho^A\pm\rho^\lambda)_{PP'}  = 0.
\end{eqnarray}
We then have unconstrained matrix elements $\delta\rho^{(1)}_{PH}$ and
\begin{subequations}
  \label{eq:56}
\begin{eqnarray}
  \label{eq:56a}
      \delta\rho^{(2)}_{PP'} &=&
                    \sum_H \delta\rho^{(1)}_{PH} \delta\rho^{(1)}_{HP'} , \\
  \label{eq:56b}
      \delta\rho^{(2)}_{HH'} &=&
                  - \sum_P \delta\rho^{(1)}_{HP} \delta\rho^{(1)}_{PH'} .
\end{eqnarray}
\end{subequations}
In summary, for {\em direct polarizations}, for which the polarizing
orbital $|\lambda\rangle$ is calculated in the {\em even} system, we
obtain the nonstandard ph structure (\ref{eq:56}) of
$\delta\rho$. However, for {\em inverse polarizations}, for which the
polarizing orbital $|\lambda\rangle$ is calculated in the {\em odd}
system, we obtain the standard ph structure (\ref{eq:57})
of $\delta\rho$.

From these considerations, it appears that a rigorous analysis of the
{\HF} polarization effects, based on the elements solely determined in the
even system, does not exist, and one must make further simplifying
assumptions. The easiest way out is to neglect the differences
between the polarizing orbitals calculated in the even and odd
systems and use equations pertaining to inverse polarizations along
with $|\lambda\rangle$ determined in the even system. In what follows, we use
this strategy.

\subsubsection{Corrections to energies}
\label{sec2.2.2}
Equations for the polarization corrections to the s.p.\
energies can be derived by comparing the
self-consistent energies in even and odd systems. Inserting
the odd-system density matrices (\ref{eq:1}) into the odd-system
energy (\ref{eq:50b}), we obtain:
\begin{eqnarray}
  \label{eq:5}
  E^{A\pm1}
  &=& E^A \pm t^{}_{\lambda\lambda} + \sum_{ii'} t^{}_{i'i} \delta\rho^{}_{ii'}
   + \thalf {\bar v}^{}_{\lambda\lambda\lambda\lambda} \nonumber \\
  &+& \thalf \sum_{ii'kk'} \delta\rho^{}_{i'i} {\bar v}^{}_{ik'i'k} \delta\rho^{}_{kk'}  \nonumber \\
  &\pm& \thalf \sum_{ii'} \rho^{A}_{i'i} {\bar v}^{}_{i\lambda i'\lambda} \pm \thalf \sum_{kk'} {\bar v}^{}_{\lambda k' \lambda k} \rho^{A}_{kk'} \nonumber \\
  &\pm& \thalf \sum_{ii'} \delta\rho_{i'i} {\bar v}^{}_{i\lambda i'\lambda} \pm \thalf \sum_{kk'} {\bar v}^{}_{\lambda k' \lambda k} \delta\rho_{kk'} \nonumber \\
  &+& \thalf \sum_{ii'kk'} \rho^{A}_{i'i} {\bar v}^{}_{ik'i'k} \delta\rho^{}_{kk'} \nonumber \\
  &+& \thalf \sum_{ii'kk'} \delta\rho^{}_{i'i} {\bar v}^{}_{ik'i'k} \rho^{A}_{kk'} .
\end{eqnarray}
We now use the following facts and definitions:
\begin{subequations}
\begin{eqnarray}
  \label{eq:55a}
 h^{A}_{i'i}     &=&  t^{}_{i'i}  + \sum_{kk'} {\bar v}^{}_{i'k'ik} \rho^{A}_{kk'} , \\
  \label{eq:55b}
  e_\lambda      &=&  h^{A}_{\lambda\lambda} \\
  \label{eq:55c}
        0        &=& {\bar v}^{}_{\lambda\lambda\lambda\lambda} , \\
  \label{eq:55d}
 h^\lambda_{i'i} &=&  {\bar v}^{}_{i'\lambda i\lambda} , \\
  \label{eq:55e}
 \delta h_{i'i}  &=& \sum_{kk'}  {\bar v}^{}_{i'k'ik}   \delta\rho^{}_{kk'} .
\end{eqnarray}
\end{subequations}
Equation (\ref{eq:55a}) is the standard definition of the {\HF} mean
field in the $A$-particle system and $e_\lambda$ (\ref{eq:55b}) is
its diagonal matrix element in the self-consistent basis. Equation
(\ref{eq:55c}) is a simple consequence of the antisymmetry of the
two-body matrix elements and represents the fact that in the {\HF}
approximation there is no self interaction (SI). Equations (\ref{eq:55d})
and (\ref{eq:55e}) define the mean-field potentials generated by the
polarizing orbital $|\lambda\rangle$ and correction $\delta\rho$,
respectively. In terms of these definitions, the odd-system energy
can be written as,
\begin{eqnarray}
  \label{eq:5b}
  E^{A\pm1}
  &=& E^A \pm e_\lambda + \sum_{ii'} h^{A}_{i'i} \delta\rho^{}_{ii'} \nonumber \\
   &&  \pm   \sum_{ii'} h^\lambda_{i'i} \delta\rho_{ii'}
  + \thalf \sum_{ii'} \delta h_{i'i} \delta\rho_{ii'}.
\end{eqnarray}

Up to now, expression (\ref{eq:5b}) is exact. To simplify it, we can
use the RPA expansion (\ref{eq:52}) and thus conditions (\ref{eq:57}),
and neglect terms beyond second order.
In the basis of particle and hole states, the mean-field Hamiltonian
$h^{A}_{i'i}$ is by definition diagonal; therefore, owing to
Eqs.~(\ref{eq:57}), the third term on the
right-hand side is of the second order in $\delta\rho^{(1)}$. Similarly, the
fifth term is obviously of the second order too. However, unless we
assume that $h^\lambda$ is small (of the first RPA
order), the fourth term may contain subleading second-order terms,
which we cannot treat. Therefore, to have a consistent
second-order expression for the energy of the $A\pm1$ system, we must
make the assumption of $h^\lambda$ being small
as compared to $h^A$. This assumption can also be understood as $\rho^\lambda$
being small as compared to $\rho^A$, that is, the system being
appropriately heavy.

In fact, such an assumption can partially be tested by keeping
the leading-order (second-order) pp$'$ and hh$'$ matrix elements of the fourth term,
which depend on the leading-order (first-order) matrix elements of $\delta\rho$.
Then, we obtain the following approximate expression,
\begin{eqnarray}
  \label{eq:5cc}
  E^{A\pm1}
  &=& E^A \pm e_\lambda + \sum_{ph} (e_p-e_h) \delta\rho^{}_{ph} \delta\rho^{}_{hp}  \nonumber \\
  &+& \thalf \sum_{ph}\delta h_{ph} \delta\rho_{hp}
   +  \thalf \sum_{ph}\delta h_{hp} \delta\rho_{ph}\nonumber \\
  &\pm&  \sum_{pp'h} h^\lambda_{p'p} \delta\rho_{ph} \delta\rho_{hp'}
   \mp   \sum_{hh'p} h^\lambda_{h'h} \delta\rho_{hp} \delta\rho_{ph'} \nonumber \\
  &\pm&  \sum_{ph}   h^\lambda_{ph}  \delta\rho_{hp}
   \pm   \sum_{ph}   h^\lambda_{hp}  \delta\rho_{ph} .
\end{eqnarray}
This can be summarized in the form of polarization corrections to
energies of odd states $\delta{}E$,
\begin{eqnarray}
  \label{eq:5da}
  E^{A\pm1}
  &=& E^A \pm e_\lambda + \delta{}E    ,
\end{eqnarray}
or polarization corrections to
s.p.\ energies $\delta{}e_\lambda$,
\begin{eqnarray}
  \label{eq:5dc}
  E^{A\pm1}
  &=& E^A \pm (e_\lambda + \delta{}e_\lambda)  ,
\end{eqnarray}
for
\begin{eqnarray}
  \label{eq:5db}
\delta{}E = \pm \delta{}e_\lambda  &=& \thalf
  \left(
    \begin{array}{cc}
      \delta\rho^*, & \delta\rho \\
    \end{array}
  \right)
  \left(
    \begin{array}{cc}
      A'   & B  \\
      B^*  & A'^* \\
    \end{array}
  \right)
  \left(
    \begin{array}{c}
      \delta\rho^{}   \\
      \delta\rho^{*}\\
    \end{array}
  \right)  \nonumber \\
&\pm&
  \left(
    \begin{array}{cc}
      \delta\rho^*, & \delta\rho \\
    \end{array}
  \right)
  \left(
    \begin{array}{c}
      h^{\lambda}   \\
      h^{\lambda*}\\
    \end{array}
  \right) ,
\end{eqnarray}
where $\delta\rho$ and $h^{\lambda}$ represent vectors of ph matrix elements,
 $\delta\rho_{ph}$ and $h^\lambda_{ph}$, respectively, that is,
\begin{subequations}
  \label{eq:e1}
\begin{eqnarray}
  \label{eq:e1a}
   h^\lambda_{ph} &=& {\bar v}^{}_{p\lambda h\lambda} , \\
  \label{eq:e1b}
   h^{\lambda*}_{ph} = h^\lambda_{hp} &=& {\bar v}^{}_{h\lambda p\lambda} ,
\end{eqnarray}
\end{subequations}
and matrices $A'$ and $B$,
\begin{subequations}
  \label{eq:e}
\begin{eqnarray}
  \label{eq:eap}
  A'_{p'h',ph} &=&  A_{p'h',ph}
               \pm  h^\lambda_{p'p}\delta_{h'h}
               \mp   h^\lambda_{hh'}\delta_{pp'}
                , \\
  \label{eq:ea}
  A_{p'h',ph} &=&  (e_p-e_h)\delta_{pp'}\delta_{hh'} +  {\bar v}^{}_{hp'ph'}, \\
  \label{eq:eb}
  B_{p'h',ph} &=& {\bar v}^{}_{pp'hh'} ,
\end{eqnarray}
\end{subequations}
build the RPA matrix
$  \left(
    \begin{array}{cc}
      A'   & B  \\
      B^*  & A'^* \\
    \end{array}
  \right)$.

We see that the second-order terms depending on $h^\lambda$, which we
have kept in Eq.~(\ref{eq:5cc}), lead to modified matrix elements $A'_{p'h',ph}$, as
compared to the standard RPA matrix $A_{p'h',ph}$. In this formulation, the RPA
equations do depend on the polarizing orbital $\lambda$. In
Sec.~\ref{sec3}, we perform numerical calculations with and without
these terms, and we check that they play a minor role and can be
safely omitted, thus supporting the validity of the assumption about
the smallness of $h^\lambda$.

\subsubsection{Equation for $\delta\rho$}
\label{sec2.2.3}
Equation for the correction $\delta\rho$ can be derived from the fact
that the density matrix of Eq.~(\ref{eq:1}) is a self-consistent solution
of the {\HF} equations in the $(A\pm1)$-particle system,
\begin{eqnarray}
  \label{eq:9}
  0 &=& \left\lbrack h^{A\pm1}, \rho^{A\pm1} \right\rbrack \nonumber\\
  &=& \left\lbrack h^{A} \pm h^\lambda +\delta h, \rho^{A} \pm \rho^\lambda + \delta\rho^{} \right\rbrack .
\end{eqnarray}
As previously, we neglect differences between the orbitals
$|\lambda\rangle$ calculated in even and odd systems, that is, we
have $\left\lbrack h^{A\pm1}, \rho^\lambda \right\rbrack=0$.
Moreover, since $\rho^{A}$ is the self-consistent solution of the
$A$-particle system, we have $\left\lbrack h^{A}, \rho^{A}
\right\rbrack =0$, which gives
\begin{eqnarray}
  \label{eq:9a}
  0 &\!\!=\!\!& \left\lbrack   h^{A}    , \delta\rho^{} \right\rbrack
      \pm \left\lbrack h^\lambda, \rho^{A} \right\rbrack
      \pm \left\lbrack h^\lambda, \delta\rho^{} \right\rbrack
      + \left\lbrack\delta h, \rho^{A} \right\rbrack
      + \left\lbrack\delta h, \delta\rho^{} \right\rbrack .  \nonumber\\
\end{eqnarray}

In the leading (first) order, the last term, quadratic in the density
$\delta\rho$, can be dropped, and we also drop the second-order matrix elements
$\delta\rho_{pp'}$ and
$\delta\rho_{hh'}$. Then, the ph and hp matrix elements of the above equation read,
\begin{subequations}
\begin{eqnarray}
  \label{eq:10}
  0 &=& \bigl( e_p - e_h \bigr) \delta\rho_{ph}
  \pm h^\lambda_{ph}
  \pm \sum_{p'} h^\lambda_{pp'} \delta\rho_{p'h}  \nonumber \\
  &&\mp \sum_{h'} \delta\rho_{ph'} h^\lambda_{h'h}
  + \delta h_{ph} \nonumber\\
    &=& (A' \delta\rho)_{ph} + (B \delta\rho^*)_{ph} \pm h^{\lambda}_{ph}, \\[2ex]
  \label{eq:10a}
  0 &=& \bigl( e_h - e_p \bigr) \delta\rho_{hp}
  \mp h^\lambda_{hp}
  \pm \sum_{h'} h^\lambda_{hh'} \delta\rho_{h'p} \nonumber \\
  &&\mp \sum_{p'} \delta\rho_{hp'} h^\lambda_{p'p}
  -\delta h_{hp} \nonumber\\
    &=& -(A'^* \delta\rho^*)_{ph} - (B^* \delta\rho)_{ph} \mp h^{\lambda*}_{ph},
\end{eqnarray}
\end{subequations}
and in the matrix notation they can be written as,
\begin{eqnarray}
  \label{eq:10b}
  \left(
    \begin{array}{cc}
      A'   & B  \\
      B^*  & A'^* \\
    \end{array}
  \right)
  \left(
    \begin{array}{c}
      \delta\rho^{}   \\
      \delta\rho^{*}\\
    \end{array}
  \right)
 = \mp
  \left(
    \begin{array}{c}
      h^{\lambda}   \\
      h^{\lambda*}\\
    \end{array}
  \right) .
\end{eqnarray}
Here again we see that the matrix elements of $h^{\lambda}$ must be at least of
the same RPA order (the first-order) as are those of $\delta\rho$.

Condition  (\ref{eq:10b}) is exactly equal to the
condition that the total energy of the odd system (\ref{eq:5da}) is
stationary with respect to correction $\delta\rho$. In other words,
vanishing variation of $\delta{}E$, Eq.~(\ref{eq:5db}), with respect to $\delta\rho$
gives Eq.~(\ref{eq:10b}).
Then, at the
stationary point, the correction to the total
energy reads
\begin{eqnarray}
  \label{eq:5f}
  \delta E
  &=&
  - \thalf
  \left(
    \begin{array}{cc}
      \delta\rho^*, & \delta\rho \\
    \end{array}
  \right)
  \left(
    \begin{array}{cc}
      A'   & B  \\
      B^*  & A'^* \\
    \end{array}
  \right)
  \left(
    \begin{array}{c}
      \delta\rho^{}   \\
      \delta\rho^{*}\\
    \end{array}
  \right) ,
\end{eqnarray}
that is, for a positive-definite RPA matrix, the correction to the total
energy is always negative,
irrespective of adding or subtracting a particle. For the corrections to s.p.\ energies
we have
\begin{eqnarray}
  \label{eq:5g}
  \delta e_\lambda
  &=& \mp\thalf
  \left(
    \begin{array}{cc}
      \delta\rho^*, & \delta\rho \\
    \end{array}
  \right)
  \left(
    \begin{array}{cc}
      A'   & B  \\
      B^* & A'^* \\
    \end{array}
  \right)
  \left(
    \begin{array}{c}
      \delta\rho^{}   \\
      \delta\rho^{*}\\
    \end{array}
  \right) ,
\end{eqnarray}
that is, particle states move down and hole states move up.
In view of Eq.~(\ref{eq:10b}), corrections (\ref{eq:5g}) can
also be written in two other equivalent forms:
\begin{eqnarray}
  \delta e_\lambda
  &=& \mp\thalf
\left(\begin{array}{cc}
h^{\lambda*}, & h^{\lambda}\end{array}\right)\left(\begin{array}{cc}
A' & B\\
B^{*} & A'^{*}
\end{array}\right)^{-1}\left(\begin{array}{c}
h^{\lambda}\\
h^{\lambda*}
\end{array}\right)\label{eq:111a} , \\
  \delta e_\lambda
  &=& -\thalf
  \left(
    \begin{array}{cc}
      \delta\rho^*, & \delta\rho \\
    \end{array}
  \right)
  \left(
    \begin{array}{c}
      h^{\lambda}   \\
      h^{\lambda*}\\
    \end{array}
  \right) . \label{eq:111b}
\end{eqnarray}

In Eq.~(\ref{eq:111a}), the inverse of the RPA matrix can be
calculated either through its eigenvectors or through the RPA
amplitudes. In the second case, we use the RPA equations and
completeness relations~\cite{[RS80]},
\begin{eqnarray}
\left(\begin{array}{cc}
A' & B\\
B^{*} & A'^{*}
\end{array}\right)\!\!
\left(\begin{array}{cc}
X & -Y^{*}\\
Y & -X^{*}
\end{array}\right)
&\!\!=\!\!&\left(\begin{array}{cc}
X & Y^{*}\\
-Y & -X^{*}
\end{array}\right)\!\!
\left(\begin{array}{cc}
\hbar\omega & 0\\
0 & \hbar\omega
\end{array}\right),
\label{eq:112a} \nonumber \\
\end{eqnarray}
\begin{eqnarray}
\left(\begin{array}{cc}
X & -Y^{*}\\
Y & -X^{*}
\end{array}\right)
\left(\begin{array}{cc}
X^+ & -Y^{+}\\
Y^T & -X^{T}
\end{array}\right)
&=&\left(\begin{array}{cc}
1 & 0\\
0 & 1
\end{array}\right)
\,,\label{eq:112b}
\end{eqnarray}
where $\hbar\omega$ is a diagonal matrix of positive RPA eigenvalues.
This gives
\begin{widetext}
\begin{equation}
\left(\begin{array}{cc}
A' & B\\
B^{*} & A'^{*}
\end{array}\right)=\left(\begin{array}{cc}
X & Y^{*}\\
-Y & -X^{*}
\end{array}\right)\left(\begin{array}{cc}
\hbar\omega & 0\\
0 & \hbar\omega
\end{array}\right)\left(\begin{array}{cc}
X^{\dagger} & -Y^{\dagger}\\
Y^{T} & -X^{T}
\end{array}\right)\,.\label{eq:112}
\end{equation}
The inverse of the RPA matrix exists if all eigenvalues are non-zero,
and has the form,
\begin{equation}
\left(\begin{array}{cc}
A' & B\\
B^{*} & A'^{*}
\end{array}\right)^{-1}=\left(\begin{array}{cc}
X & -Y^{*}\\
Y & -X^{*}
\end{array}\right)\left(\begin{array}{cc}
\hbar\omega^{-1} & 0\\
0 & \hbar\omega^{-1}
\end{array}\right)\left(\begin{array}{cc}
X^{\dagger} & Y^{\dagger}\\
-Y^{T} & -X^{T}
\end{array}\right).\label{eq:113}
\end{equation}
\end{widetext}
Finally, in terms of the RPA amplitudes and energies, corrections~(\ref{eq:111a})
then become equal to~\cite{[Bla86]},
\begin{eqnarray}
  \delta e_\lambda
  &=&  \mp\sum_{\omega>0}\frac{\left\vert \sum_{ph}h^{\lambda*}_{ph}X_{ph}^{\omega}+h^{\lambda}_{ph}Y_{ph}^{\omega}\right\vert ^{2}}{\hbar\omega}
\label{eq:117} .
\end{eqnarray}

\subsection{Polarization corrections for density functionals}
\label{sec2.4}

Let us now rederive the polarization corrections of Sec.~\ref{sec2.2}
for the case of the total energy given by a minimization of an EDF, and
not of the {\HF} average of a Hamiltonian. Differences between these
two cases can be of dual origin. First, a quasilocal EDF built as the
most general quadratic function of densities deviates from a {\HF}
average of a zero-range momentum-dependent interaction unless its
coupling constants obey a specific set of linear conditions, see,
e.g., Refs.~\cite{[Dob10],[Cha10]}. For the Skyrme EDF, these
conditions can be formulated as a linear dependence of the time-odd
coupling constants on the time-even ones, and a linear
dependence between the isovector and isoscalar spin-orbit coupling
constants~\cite{[Rei95]}. In this work we only consider
EDFs of this type. The second reason for differences arises because of
so-called density-dependent
interactions, which also lead to EDFs that are not equal to {\HF} averages
of Hamiltonians.

Focusing on this second case, we now consider EDFs determined by the
{\HF} averaging of antisymmetrized density-dependent matrix elements
${\bar{v}}_{i'k'ik}[\rho]$. Then, the total energies read,
\begin{subequations}
  \label{eq:60}
\begin{eqnarray}
  \label{eq:60a}
  E^{A} &=& {\rm Tr}(t\rho^{A}) +
  \thalf {\Tr}_1{\Tr}_2(\rho^{A}\bar{v}[\rho^{A}]\rho^{A}) , \\
  \label{eq:60b}
  E^{A\pm1} &=& {\rm Tr}(t\rho^{A\pm1}) \nonumber \\
   &&+\thalf {\Tr}_1{\Tr}_2(\rho^{A\pm1}\bar{v}[\rho^{A\pm1}]\rho^{A\pm1}) .
\end{eqnarray}
\end{subequations}
We see that both energies, apart from the standard quadratic dependencies
on densities, cf.~Eqs.~(\ref{eq:50}), do depend on densities through
the density dependence of interactions. These latter dependence
precludes comparing energies of even and odd systems, unless we
make the simplifying assumption that $\bar{v}[\rho^{A\pm1}]$
and $\bar{v}[\rho^{A}]$ can be connected by a second-order expansion
in $\rho^{A\pm1}-\rho^{A}$. From Eq.~(\ref{eq:1}) we see again that
this requires $\rho^\lambda$ to be of the same (first) RPA order
as $\delta{\rho}$. Under this assumption, we have
\begin{eqnarray}
  \label{eq:61}
 && \!\!\bar{v}_{i'k'ik}[\rho^{A\pm1}] \simeq \bar{v}_{i'k'ik}[\rho^{A}]
  \pm \sum_{mn}\frac{\partial \bar{v}_{i'k'ik}}{\partial\rho_{mn}}
   \left(\rho^\lambda_{mn}\pm\delta{\rho}_{mn}\right) \nonumber \\
 && \!\!+\thalf\!\!\!\! \sum_{m'n'mn}\!\!\frac{\partial^2 \bar{v}_{i'k'ik}}{\partial\rho_{mn}\partial\rho_{m'n'}}
   \left(\rho^\lambda_{mn}  \pm\delta{\rho}_{mn}\right)
   \left(\rho^\lambda_{m'n'}\pm\delta{\rho}_{m'n'}\right) ,  \nonumber \\
\end{eqnarray}
where all partial derivatives must be evaluated at $\rho\equiv\rho^{A}$.

We can now insert Eqs.~(\ref{eq:1}) and (\ref{eq:61}) into the odd-system
energy (\ref{eq:60b}) and obtain up to the second order in
$\pm\rho^\lambda + \delta{\rho}$,
\begin{eqnarray}
  \label{eq:62}
  E^{A\pm1} &=& E^{A} + {\rm Tr}\tilde{h}^A(\pm\rho^\lambda + \delta{\rho}) \nonumber \\
   &&+\thalf {\Tr}_1{\Tr}_2(\pm\rho^\lambda + \delta{\rho})\tilde{\tilde{v}}
                          (\pm\rho^\lambda + \delta{\rho}) ,
\end{eqnarray}
where the mean-field Hamiltonian $\tilde{h}^A$,
\begin{eqnarray}
  \label{eq:63}
 \tilde{h}^{A}_{i'i}     &=&  t^{}_{i'i}  + \sum_{kk'} \tilde{v}^{}_{i'k'ik} \rho^{A}_{kk'} ,
\end{eqnarray}
and effective two-body
matrix elements, $\tilde{v}_{i'k'ik}$ and $\tilde{\tilde{v}}_{i'k'ik}$,
contain rearrangement terms,
\begin{subequations}
  \label{eq:64}
\begin{eqnarray}
  \label{eq:64a}
  \tilde{v}_{i'k'ik} &=& \bar{v}_{i'k'ik}
  +\thalf \sum_{j'j}\frac{\partial \bar{v}_{j'k'jk}}{\partial\rho_{ii'}}\rho^{A}_{jj'} , \\
  \label{eq:64b}
  \tilde{\tilde{v}}_{i'k'ik} &=& \bar{v}_{i'k'ik}
  + \sum_{j'j}\left(\frac{\partial \bar{v}_{j'k'jk}}{\partial\rho_{ii'}}
                  + \frac{\partial \bar{v}_{j'i'ji}}{\partial\rho_{kk'}}
                                                           \right)\rho^{A}_{jj'}
\nonumber \\
  &&+\thalf \sum_{j'm'jm}\frac{\partial^2 \bar{v}_{j'm'jm}}
                           {\partial\rho_{ii'}\partial\rho_{kk'}}
   \rho^{A}_{jj'}  \rho^{A}_{mm'} .
\end{eqnarray}
\end{subequations}
The redefined two-body matrix elements allow us to write the
odd-system energy in the form analogous to Eq.~(\ref{eq:5b}),
\begin{eqnarray}
  \label{eq:65}
  E^{A\pm1}
  &=& E^A \pm e_\lambda + \sum_{ii'} \tilde{h}^{A}_{i'i} \delta\rho^{}_{ii'}
  + \thalf\tilde{\tilde{h}}^\lambda_{\lambda\lambda}\nonumber \\
   &&  \pm   \sum_{ii'} \tilde{\tilde{h}}^\lambda_{i'i} \delta\rho_{ii'}
  + \thalf \sum_{ii'} \delta \tilde{\tilde{h}}_{i'i} \delta\rho_{ii'},
\end{eqnarray}
but with the following redefinitions,
\begin{subequations}
\begin{eqnarray}
  \label{eq:66b}
  e_\lambda      &=&  \tilde{h}^{A}_{\lambda\lambda} \\
  \label{eq:66c}
   \tilde{\tilde{h}}^\lambda_{\lambda\lambda} &=& \tilde{\tilde{v}}_{\lambda\lambda\lambda\lambda} , \\
  \label{eq:66d}
 \tilde{\tilde{h}}^\lambda_{i'i} &=&  \tilde{\tilde{v}}_{i'\lambda i\lambda} , \\
  \label{eq:66e}
 \delta \tilde{\tilde{h}}_{i'i}  &=& \sum_{kk'} \tilde{\tilde{v}}_{i'k'ik}   \delta\rho^{}_{kk'} .
\end{eqnarray}
\end{subequations}
We see that the first order rearrangement terms (\ref{eq:64a}) become
fully absorbed in the s.p.\ energies, which are now, as
usual, the eigenvalues of mean fields $\tilde{h}^{A}$. Moreover, both
the polarizing fields $\tilde{\tilde{h}}^\lambda$ and RPA matrices
$A$ and $B$, see Eqs.~(\ref{eq:e1}) and~(\ref{eq:e}),
must now be determined by using the second-order
rearrangement terms (\ref{eq:64b}). Therefore, owing to the fact
that the effective two-body matrix elements (\ref{eq:64a}) are not
antisymmetric, the SI term (\ref{eq:66c}),
\begin{eqnarray}
  \label{eq:88}
E^\lambda_{\text{SI}} &=& \thalf\tilde{\tilde{h}}^\lambda_{\lambda\lambda} ,
\end{eqnarray}
is non-zero, and
explicitly appears in Eq.~(\ref{eq:65}). This leads to corrections
to s.p.\ energies now having the form,
\begin{eqnarray}
  \label{eq:68}
  \delta e_\lambda = \pm \delta E
  &=&  \pm \left( \delta E^\lambda_{\text{SIF}}
         +               E^\lambda_{\text{SI}} \right) ,
\end{eqnarray}
where, based on the analogy with Eq.~(\ref{eq:117}),
the first term can be called self-interaction-free (SIF) polarization correction,
\begin{eqnarray}
  \label{eq:68a}
 \delta  E^\lambda_{\text{SIF}}
  &=&  -\sum_{\omega>0}\frac{\left\vert \sum_{ph}\tilde{\tilde{h}}^{\lambda*}_{ph}
                                   X_{ph}^{\omega}+\tilde{\tilde{h}}^{\lambda}_{ph}
                                   Y_{ph}^{\omega}\right\vert ^{2}}{\hbar\omega} .
\end{eqnarray}

The second-order mean fields
$\tilde{\tilde{h}}^\lambda_{i'i}$ (\ref{eq:66d}) and
$\delta \tilde{\tilde{h}}_{i'i}$ (\ref{eq:66e}) are simply
related to the linearized first-order mean fields, that is,
\begin{subequations}
\begin{eqnarray}
  \label{eq:67d}
 \tilde{\tilde{h}}^\lambda_{i'i} &=&
    \sum_{k'k}\frac{\partial\tilde{h}_{i'i}}
                  {\partial\rho_{k'k}}\raisebox{-1em}{\rule{0.02em}{2.5em}}_{\,\rho=\rho^A}\,\rho^\lambda_{k'k} , \\
  \label{eq:67e}
 \delta \tilde{\tilde{h}}_{i'i}  &=&
    \sum_{k'k}\frac{\partial\tilde{h}_{i'i}}
                  {\partial\rho_{k'k}}\raisebox{-1em}{\rule{0.02em}{2.5em}}_{\,\rho=\rho^A}\,\delta\rho_{k'k} .
\end{eqnarray}
\end{subequations}
These expressions can be explicitly verified directly from definitions
(\ref{eq:64}). They are extremely useful in practical applications,
because: (i) the second-order mean-fields (\ref{eq:67d}) that define
the polarization vertex (\ref{eq:68a}) can be determined without
explicitly calculating the second derivatives of matrix elements,
(ii) the amplitude mean-fields (\ref{eq:67e}) are the only objects
that one has to calculate when using the iterative methods to solve
the RPA equations~\cite{[Toi10]}, and (iii) exactly the same piece of
code can be used to calculate both mean fields (\ref{eq:67d}) and
(\ref{eq:67e}).

\subsubsection{The self interaction}
\label{sec2.4.1}
The SI term
(\ref{eq:88}), where a particle interacts with the mean field
generated by itself, is unphysical, because in reality each nucleon
should interact with the other nucleons only.  As discussed in
Sec.~\ref{sec2.2.2}, for an EDF generated by Hamiltonian, no SI
appears. On the other hand, EDFs generated by density-dependent
interactions do produce the SI.

An EDF has a one-body SI if it gives non-zero energy for a single
nucleon state.  This was discussed in Ref.~\cite{[Cha10]}, where it
has been shown how the one-body SI of a Skyrme EDF can be removed by
introducing extra constraints between the Skyrme coupling constants.
In general, the SI results from the violation of the antisymmetry of
effective matrix elements (\ref{eq:64b}). For density functionals
used in electronic structure calculations, an SI correction
was originally introduced by Perdew and
Zunger~\cite{[Per81]}, and numerous variations and improvements of the
method have been later constructed. A short review of the various SI-correction
methods used in electronic structure calculations can be found in
Ref.~\cite{[Leg02]}. In nuclear physics context, the SI problem in
connection with density-dependent Skyrme interactions was early on
discussed in Ref.~\cite{[Str78]} and more recently in
Ref.~\cite{[Klu09]}. In Refs.~\cite{[Lac09],[Ben09]}, the SI problem
and ways to correct it was discussed in detail, together with the
related concept of self pairing, see also Sec.~\ref{sec2.3}.

Within the Skyrme EDF approach without pairing, the SI results from the zero-range
density-dependent interaction and from the Coulomb exchange, which is
treated in the Slater approximation. Since in the polarization
correction (\ref{eq:68}) the SI term appears explicitly, one can
simply remove it from this expression and thus obtain SIF
result (\ref{eq:68a}). However, we stress here that the self-consistent calculations
performed in odd nuclei do contain the SI term, and lead to the
polarization correction (\ref{eq:68}) with the SI term included.

\subsection{Polarization corrections with pairing}
\label{sec2.3}

For a paired system corresponding to the average number of particles $A$, one
diagonalizes the quasiparticle Hamiltonian $\cH^A$, which in the
standard double-dimension representation~\cite{[RS80]} reads,
\begin{equation}
\cH^A = \cT + \cG^A -\lambda\cN = \begin{pmatrix} h^A-\lambda & \Delta^A  \\ -\Delta^{A*} &
-h^{A*}+\lambda \end{pmatrix} ,
\label{H-dens}
\end{equation}
where
\begin{equation}
\cT = \begin{pmatrix} t & 0  \\
                      0 & -t^* \end{pmatrix} , ~~
\cG^A = \begin{pmatrix} \Gamma^A & \Delta^A  \\
                    -\Delta^{A*} &   -\Gamma^{A*} \end{pmatrix} ,  ~~
\cN = \begin{pmatrix} 1 & 0  \\
                      0 & -1 \end{pmatrix} ,
\label{H-pot}
\end{equation}
and where $\Gamma^A_{i'i}$ and $\Delta^A_{i'k'}$ are the ph and pp mean
potentials, respectively. In what follows, for clarity we write only
one Fermi energy $\lambda$ -- generalizations to separate neutron and proton
Fermi energies being obvious.

The eigenequation for $\cH^A$ defines one-quasiparticle
states $\cU_L$ and one-quasiparticle energies $E_L$,
\begin{equation}
\label{eq:79}
\cH^A \cU_L= E_L \cU_L ,
\end{equation}
where positive (negative) indices $L>0$  ($L<0$) correspond to
positive (negative) quasiparticle energies $E_L>0$
$(E_L<0)$ of quasiparticle (quasihole) states.
Then, the basic dynamical quantity describing the system is
the generalized density matrix, $(\cR^A)^2=\cR^A$,
\begin{equation}
\cR^A = \sum_{L<0} \cU_L\cU_L^+
      = \begin{pmatrix} \rho^A & \kappa^A  \\ -\kappa^{A*} & 1-\rho^{A*} \end{pmatrix},
\label{eq:70}
\end{equation}
which projects states on the space of occupied quasihole states,
\begin{equation}
\cR^A \cU_L = \left\{\begin{array}{lr} 0 &    \quad \mbox{for $L>0$}, \\
                                   \cU_L &    \quad \mbox{for $L<0$},
                     \end{array}\right.\vspace*{3mm}
\end{equation}
and depends on the density matrix $\rho^A$ and pairing tensor $\kappa^A$.
When the quasiparticle and quasihole states are arranged as columns
of matrix $\cU$ in doubled dimensions, they form the matrix of the
Bogoliubov transformation~\cite{[RS80]},
\begin{equation}
\cU = \left(\begin{array}{cc} U & V^*  \\ V & U^* \end{array}\right) ,
\label{HFB-eq}
\end{equation}
in terms of which we have $\rho^A = V^* V^T$ and $\kappa^A = V^* U^T$.

The generalized density matrix of an odd system $\cR^{A'}$
is obtained by the blocking procedure~\cite{[RS80],[Ber09]},
whereupon one occupied quasihole state for $L=-\Lambda<0$ is replaced
by its empty quasiparticle partner for $L=\Lambda>0$.
Then, Eqs.~(\ref{H-dens})--(\ref{eq:70}) are solved self-consistently again,
and the QRPA polarization correction $\delta\cR$ is defined
in analogy with Eq.~(\ref{eq:1}) as,
\begin{equation}
\cR^{A'}  = \cR^A + \cR^\Lambda + \delta\cR   ,
\label{eq:73}
\end{equation}
where the generalized density matrix $\cR^\Lambda$ describes the
blocked orbital [see Eq.~(\ref{eq:70})],
\begin{equation}
\label{eq:82}
\cR^\Lambda  =    - \cU_{-\Lambda} \cU_{-\Lambda}^+
                  + \cU_{ \Lambda} \cU_{ \Lambda}^+ .
\end{equation}
This gives the density matrix and pairing tensor of the odd
system in the form,
\begin{subequations}
\label{eq:kappdel2}
\begin{eqnarray}
\rho^{A'}  =\rho^{A}   + \rho^\Lambda   + \delta\rho , \\
\kappa^{A'}=\kappa^{A} + \kappa^\Lambda + \delta\kappa ,
\end{eqnarray}
\end{subequations}
with explicit contributions coming from the blocked orbital given by,
\begin{subequations}
\label{eq:kappdel3}
\begin{eqnarray}
\rho_{\alpha\beta}^\Lambda  =-V_{\alpha\Lambda}^{*}V_{\beta\Lambda}+U_{\alpha\Lambda}U_{\beta\Lambda}^{*} , \\
\kappa_{\alpha\beta}^\Lambda=-V_{\alpha\Lambda}^{*}U_{\beta\Lambda}+U_{\alpha\Lambda}V_{\beta\Lambda}^{*}  .
\end{eqnarray}
\end{subequations}

Density matrices and pairing tensors, self-consistently calculated in
even and odd systems, determine the corresponding total energies with pairing as,
\begin{subequations}
  \label{eq:71}
\begin{eqnarray}
  \label{eq:71a}
  E^{A} &=& {\Tr}\,t\rho^{A} +
      \thalf {\Tr}\left(\Gamma^A \rho^{A} - \Delta^A\kappa^{A*}\right) \nonumber \\
        &=& \thalf {\Tr}\,\cT\cQ^{A} + \tfrac{1}{4} {\Tr}\cG^A \cQ^{A} , \\
  \label{eq:71b}
  E^{A'} &=& {\Tr}\,t\rho^{A'} +
      \thalf {\Tr}\left(\Gamma^{A'} \rho^{A'} - \Delta^{A'}\kappa^{A'*}\right) \nonumber \\
        &=& \thalf {\Tr}\,\cT\cQ^{A'} + \tfrac{1}{4} {\Tr}\cG^{A'} \cQ^{A'} ,
\end{eqnarray}
\end{subequations}
where
$\cQ^{A }\equiv\cR^{A }-\left(\begin{array}{cc} 0 & 0 \\ 0 & 1\end{array}\right)$
and
$\cQ^{A'}\equiv\cR^{A'}-\left(\begin{array}{cc} 0 & 0 \\ 0 & 1\end{array}\right)$.

Let us now discuss the case of the ph and pp potentials being determined
by averaging the corresponding ph and pp two-body antisymmetric interaction
matrix elements, that is,
\begin{subequations}
  \label{eq:72}
\begin{eqnarray}
  \label{eq:72a}
\Gamma^A_{i'i}=\sum_{k'k}{\bar{v}}^{\text{ph}}_{i'k'ik}\rho^A_{kk'} ,\\
  \label{eq:72b}
\Delta^A_{i'k'}=\thalf\sum_{ik}{\bar{v}}^{\text{pp}}_{i'k'ik}\kappa^A_{ik},
\end{eqnarray}
\end{subequations}
with the analogous equations defining the potentials in the odd
system, $\Gamma^{A'}$ and $\Delta^{A'}$. In most nuclear-physics
applications, the ph and pp matrix elements are different, which
means that the total energies do not, strictly speaking, correspond
to an average value of a Hamiltonian.

To discuss the structure of the resulting expressions, we first
analyze the situation of these matrix elements being
density-independent, that is, with the rearrangement terms ignored.
Then, the potentials depend linearly on densities, and each term in
the generalized density matrix $\cR^{A'}$ (\ref{eq:73})
gives the corresponding term in the generalized potential $\cG^{A'}$,
\begin{eqnarray}
\label{eq:74a}
\cG^{A'}  &=& \cG^A + \cG^\Lambda + \delta\cG  .
\end{eqnarray}

By inserting Eqs.~(\ref{eq:73}) and (\ref{eq:74a}) into expression
for the total energy of the odd system (\ref{eq:71b}), we easily obtain the
analogue of Eq.~(\ref{eq:5b}) with pairing, that is,
\begin{eqnarray}
  \label{eq:75}
  E^{A'}
  &=& E^A + \Tr \Big[\thalf(\cH^A+\lambda\cN)\cR^\Lambda
                     +\thalf(\cH^A+\lambda\cN)\delta\cR     \nonumber \\
  & &~~~~~~~~~       +\tfrac{1}{4}\cH^\Lambda\cR^\Lambda
                     +\thalf\cH^\Lambda\delta\cR
                     +\tfrac{1}{4}\delta\cH\delta\cR \Big],
\end{eqnarray}
where, to keep notation consistent with the unpaired case of
Sec.~\ref{sec2.2.2}, we have denoted $\cH^\Lambda\equiv\cG^\Lambda$
and $\delta\cH\equiv\delta\cG$.

At this point, we have to recall that the average number of particles in the
blocked state, $A'=\Tr\rho^{A'}=A+\Delta{A}$ for
\begin{eqnarray}
  \label{eq:76}
\Delta{A} &=& \Tr\,\rho^\Lambda + \Tr\,\delta\rho
           =  \thalf\Tr\,\cN\cR^\Lambda + \thalf\Tr\,\cN\delta\cR,
\end{eqnarray}
is not necessarily equal to $A\pm1$. Therefore, to calculate the average energy of the
$A\pm1$ system, one has to introduce a linear correction~\cite{[RS80]},
such that
\begin{eqnarray}
E^{A\pm1} &=& E^{A'} + \lambda'(A\pm1 - A') \nonumber \\
          &=& E^{A'} \pm\lambda' - \lambda'\Delta{A}  ,
\end{eqnarray}
where by definition $\lambda'= \frac{d E^{A'}}{d A'}$ is the
Fermi energy of the odd system. We then see that, under the assumption
of self-consistent Fermi energies in the even and odd systems being equal,
$\lambda'\simeq\lambda$, we have
\begin{eqnarray}
  \label{eq:77}
  E^{A\pm1}
  &=& E^A \pm\lambda + \Tr \Big[\thalf\cH^A\cR^\Lambda
                     +\thalf\cH^A\delta\cR     \nonumber \\
  & &~~~~~       +\tfrac{1}{4}\cH^\Lambda\cR^\Lambda
                     +\thalf\cH^\Lambda\delta\cR
                     +\tfrac{1}{4}\delta\cH\delta\cR \Big].
\end{eqnarray}

From this point on, derivations proceed exactly as in the case
of no pairing, Sec.~\ref{sec2.2}, so we only repeat
principal definitions and results. We assume that the blocked
quasiparticle wave function $\cU_\Lambda$, which is determined in the
odd system, is identical to that determined in the even system. Only
under such an assumption we have
$\thalf\Tr\,\cH^A\cR^\Lambda=E_\Lambda$ [see Eqs.(\ref{eq:79}) and
(\ref{eq:82})] and the analogue of Eq.~(\ref{eq:53a}) holds,
\begin{eqnarray}
  \label{eq:78}
      \delta\cR &=& \cR^A\delta\cR + \delta\cR\cR^A .
\end{eqnarray}

In Eq.~(\ref{eq:77}) we identify the SI term, analogous to that
derived without pairing (\ref{eq:88}), namely,
\begin{eqnarray}
  \label{eq:89}
E^\Lambda_{\text{SI}} &=&  \tfrac{1}{4}\Tr\,\cH^\Lambda\cR^\Lambda
                       =   \tfrac{1}{2}\Tr\,\left(\Gamma^\Lambda\rho^\Lambda
                                                 -\Delta^\Lambda\kappa^{\Lambda*}\right) ,
\end{eqnarray}
where $\Gamma^\Lambda$ and $\Delta^\Lambda$ are the mean fields generated
by the blocked quasiparticle,
\begin{subequations}
  \label{eq:92}
\begin{eqnarray}
  \label{eq:92a}
\Gamma^\Lambda_{i'i}=\sum_{k'k}{\bar{v}}^{\text{ph}}_{i'k'ik}\rho^\Lambda_{kk'} ,\\
  \label{eq:92b}
\Delta^\Lambda_{i'k'}=\thalf\sum_{ik}{\bar{v}}^{\text{pp}}_{i'k'ik}\kappa^\Lambda_{ik}.
\end{eqnarray}
\end{subequations}
Thus the SI term corresponds to the blocked quasiparticle $\Lambda$ that
interacts with the generalized mean-field potential it has generated.
By combining Eqs.~(\ref{eq:kappdel3}) and (\ref{eq:92}), we
can easily derive that
\begin{eqnarray}
  \label{eq:90}
E^\Lambda_{\text{SI}} &=& \sum_{i'k'ik}U^*_{k'\Lambda}V_  {i'\Lambda}
              \left({\bar{v}}^{\text{pp}}_{i'k'ik} - {\bar{v}}^{\text{ph}}_{i'k'ik} \right)
                               U_  {k \Lambda}V^*_{i \Lambda} ,
\end{eqnarray}
where the antisymmetry of matrix elements ${\bar{v}}^{\text{pp}}_{i'k'ik}$
and ${\bar{v}}^{\text{ph}}_{i'k'ik}$ was used.
We explicitly see that a non-zero value of $E^\Lambda_{\text{SI}}$ can
only appear when the pp and ph interactions, which define the EDF
with pairing, are not identical to one another.

Another assumption we have to make is that the mean field $\cH^\Lambda$,
generated by the blocked quasiparticle $\Lambda$, is appropriately
small -- of the first order in QRPA.
Then, the polarization corrections to paired
energies of odd states $\delta{}E$ and those to
quasiparticle energies $\delta{}E_\Lambda$ [cf.~Eqs.~(\ref{eq:5da}) and (\ref{eq:5dc})],
which are defined by
\begin{eqnarray}
  \label{eq:80}
  E^{A\pm1}
  &=& E^A \pm\lambda + E_\Lambda + \delta{}E  \nonumber \\
  &=& E^A \pm\lambda + (E_\Lambda + \delta{}E_\Lambda)  ,
\end{eqnarray}
can be expressed in the form
\begin{eqnarray}
  \label{eq:81}
\delta{}E = \delta{}E_\Lambda  &=& \thalf
  \left(
    \begin{array}{cc}
      Z^*, & Z \\
    \end{array}
  \right)
  \left(
    \begin{array}{cc}
      A   & B  \\
      B^* & A^* \\
    \end{array}
  \right)
  \left(
    \begin{array}{c}
      Z   \\
      Z^{*}\\
    \end{array}
  \right) \nonumber \\
&+&
  \left(
    \begin{array}{cc}
      Z^*, & Z \\
    \end{array}
  \right)
  \left(
    \begin{array}{c}
      W^{\Lambda}   \\
      W^{\Lambda*}   \\
    \end{array}
  \right) + E^\Lambda_{\text{SI}} ,
\end{eqnarray}
where $Z$ and $W^{\Lambda}$ represent vectors of
quasiparticle-quasihole matrix elements of $\delta\cR$ and
$\cH^\Lambda$, respectively, that is
\begin{eqnarray}
  \label{eq:84a}
   Z_{LL'}            &=& \cU_L^+ \delta\cR   \cU_{-L'},  \\
  \label{eq:84b}
    W^{\Lambda}_{LL'} &=& \cU_L^+ \cH^\Lambda \cU_{-L'},
\end{eqnarray}
for $L>L'>0$,
and $A$ and $B$ are the standard components of the QRPA matrix~\cite{[RS80]}.

Equation for $Z$ can easily be derived by following the steps presented in
Sec.~\ref{sec2.2.3} -- it simply results from the requirement that
the correction to the energy (\ref{eq:81}) is stationary, which gives,
\begin{eqnarray}
  \label{eq:85}
  \left(
    \begin{array}{cc}
      A   & B  \\
      B^* & A^* \\
    \end{array}
  \right)
  \left(
    \begin{array}{c}
      Z   \\
      Z^{*}\\
    \end{array}
  \right)
&=& - \left(
    \begin{array}{c}
      W^{\Lambda}   \\
      W^{\Lambda*}   \\
    \end{array}
  \right)
\end{eqnarray}
and
\begin{eqnarray}
  \label{eq:86}
\delta{}E  &=&  \delta E^\Lambda_{\text{SIF}} + E^\Lambda_{\text{SI}} ,
\end{eqnarray}
for
\begin{eqnarray}
  \label{eq:86f}
\delta E^\Lambda_{\text{SIF}}  &\!\!=\!\!& - \thalf
  \left(\!\!
    \begin{array}{cc}
       W^{\Lambda*}, &  W^{\Lambda} \\
    \end{array}
 \!\! \right)
  \!\!\left(\!\!
    \begin{array}{cc}
      A   & B  \\
      B^* & A^* \\
    \end{array}
 \!\! \right)^{-1}\!\!
  \left(\!\!
    \begin{array}{c}
       W^{\Lambda} \\
      W^{\Lambda*} \\
    \end{array}
  \!\!\right)  .
\end{eqnarray}
The QRPA SIF polarization correction to quasiparticle energy explicitly reads
\begin{eqnarray}
  \label{eq:86b} \!\!\!\! \!\!\!\!
 \delta E^{\Lambda}_{\text{SIF}}
  &\!\!=\!\!&   -\sum_{\omega>0}\frac{\left\vert
   \sum_{L>L'}\left(W^{\Lambda*}_{LL'}X_{LL'}^{\omega}
                   +W^{\Lambda }_{LL'}Y_{LL'}^{\omega}
              \right)\right\vert ^{2}}{\hbar\omega} ,
\end{eqnarray}
where $X_{LL'}^{\omega}$ and $Y_{LL'}^{\omega}$ are the standard QRPA
amplitudes.

\subsection{A few remarks to Sec.~\protect\ref{sec2}}
\label{sec2.5}

Before discussing numerical results in the next section, let us
briefly touch upon the problem of conserved symmetries. No specific
conserved symmetry was, in fact, assumed in the derivations presented
so far. In Appendix~\ref{sec5}, we discuss in detail implications of
conserving the spherical symmetry. For the conserved parity, the
s.p.\ wave function $\psi_\lambda(\bm{r})$ of a polarizing orbital
$\lambda$ has a definite parity $\pi_\lambda$, that is,
$\psi_\lambda(-\bm{r})=\pi_\lambda\psi_\lambda(\bm{r})$. Therefore,
the density matrix $\rho^\lambda$, which in the space
coordinates reads
$\rho^\lambda(\bm{r},\bm{r}')=\psi_\lambda(\bm{r})\psi_\lambda(\bm{r}')^*$,
is {\em always} parity even, irrespective of the
parity of the polarizing orbital:
$\rho^\lambda(-\bm{r},-\bm{r}')=\pi_\lambda^2\psi_\lambda(\bm{r})\psi_\lambda(\bm{r}')^*
=\rho^\lambda(\bm{r},\bm{r}')$. With pairing correlations included,
analogous arguments hold for quasiparticle wavefunction
$\cU_{-\Lambda}$ and generalized density matrix $\cR^\Lambda$.
Since interactions are parity-invariant, the positive parity of
$\rho^\lambda$ or  $\cR^\Lambda$ implies positive parity of mean
fields $h^\lambda$ or  $W^\Lambda$, and thus only positive-parity
phonons contribute to the energy shifts in Eqs.~(\ref{eq:68a}) or
(\ref{eq:86b}).

This means that, for conserved spherical and parity symmetries,
all polarization corrections discussed in Sec.~\ref{sec2} relate to
the RPA and QRPA channels and phonons $J^\pi$ with positive parity
$\pi=+1$ only. Therefore, the discussed polarization
corrections cannot, and do not, involve
any couplings to negative parity channels, including those to the
very important $3^-$ channel. The latter can only be treated within
the PVC methods~\cite{(Col10)}, which will be discussed in the forthcoming
study~\cite{[Tar13]}. The lack of couplings to negative-parity
channels constitutes the main drawback of the mean-field methods
in describing states in odd nuclei.

We note here that the polarization corrections studied in this work
are equivalent to the ``diagonal'' approximation to the PVC,
whereupon the polarization vertex is limited to the same state as the
one for which the PVC is evaluated. This shows again that the
parity-conservation in the vertex excludes coupling to negative-parity
phonons. Such coupling is only possible when the ``non-diagonal'' PVC
is calculated for intermediate states that involve opposite-parity s.p.\
or quasiparticle states.

There is another hypothetical possibility of including the coupling to
negative-parity phonons, namely, through a dynamical parity-breaking
of the mean field. This would require performing
generator-coordinate-like calculations based on mixing
octupole-deformed states for odd nuclei. It is, however, unclear if such an
approach can be equivalent to the PVC that includes negative-parity phonons.

Another drawback of the mean-field approach, clearly identified in
Secs.~\ref{sec2.4} and~\ref{sec2.3}, is the presence of the SI terms
in the mean-field binding energies of odd nuclei. Based on the
analyses performed within the RPA and QRPA methods, we could
explicitly identify these terms, which allows for calculating them
{\em after variation}. The explicit identification may also allow us
to remove them {\em before variation}, which will be the subject of
subsequent studies. Of course, although not explicitly ``visible'',
the SI terms are also present in the mean-field binding energies of
even nuclei, and in the future new functionals with these terms
removed should also be studied.

Finally, we note that expression (\ref{eq:90}) for the SI energy with
pairing is valid only for density-independent interactions. However,
expression (\ref{eq:89}) does take effects of density-dependent
interactions into account, provided it is evaluated for mean-field
potentials with rearrangement terms included, as derived in
Sec.~\ref{sec2.2}, that is,
\begin{eqnarray}
  \label{eq:89a}
E^\Lambda_{\text{SI}} &=&  \tfrac{1}{2}\Tr\,\left(\tilde{\tilde{h}}^\Lambda\rho^\Lambda
                                                 -\Delta^\Lambda\kappa^{\Lambda*}\right) .
\end{eqnarray}

\section{Results}
\label{sec3}
All calculations presented in this section aim at comparing
self-consistent results obtained by using the deformed solver
{\sc{hfodd}} (v2.52k)~\cite{[Sch12]} with RPA and QRPA solutions
implemented in the spherical solver {\sc{hosphe}}~\cite{[Car13]}.
We used the configuration space that includes all harmonic-oscillator
shells up to $N_0=15$

\subsection{RPA calculations in {\boldmath$^{100}$Sn\unboldmath} for the Skyrme EDF SV}
\label{sec3.1}

We begin the presentation by showing examples of calculations
performed for the case of an exact {\HF} approximation, as discussed
in Sec.~\ref{sec2.2}. To this end, we employed the
density-independent Skyrme interaction SV~\cite{[Bei75]} and we
analyzed results only for neutrons, so as to avoid effects of
density-dependent Slater approximation for the Coulomb exchange term.
On the one hand, to treat the EDF SV as fully generated by an
interaction, we included in the functional all tensor terms, which
were originally neglected~\cite{[Bei75]}. Also the ``native'' time-odd
terms of SV were included. On the other hand, as mentioned in
Sec.~\ref{sec2.2}, we neglected the so-called center-of-mass
correction to the kinetic energy.

\begin{figure}
\includegraphics[width=0.4\textwidth]{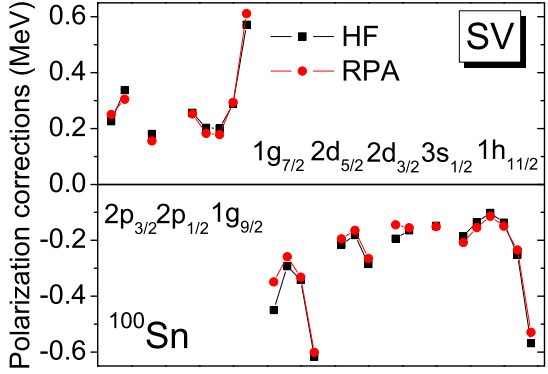}
\caption{(Color online) Comparison of polarization corrections of
selected orbitals in $^{100}$Sn, determined by using the {\HF} and RPA
methods and Skyrme EDF SV~\protect\cite{[Bei75]}, see text. Lines connect the values
obtained for different projections of the angular momentum
$|m_\lambda|=\thalf,\ldots,j_\lambda$ (from left to right).
}
\label{fig:three}
\end{figure}
In Fig.~\ref{fig:three} we test Eq.~(\ref{eq:5dc}), that is, we
compare polarization corrections,
\begin{eqnarray}
  \label{eq:48}
\delta{}e_\lambda=\pm(E^{A\pm1}-E^A)-e_\lambda ,
\end{eqnarray}
obtained from the {\HF}
energies of odd and even systems, $E^{A\pm1}$ and $E^A$, and {\HF} s.p.\
energies, $e_\lambda$, with those determined form the RPA solutions,
Eq.~(\ref{eq:117}). Apart from a few cases, the obtained
agreement is nearly perfect. This result is particularly gratifying,
because it confirms not only the analytical derivations presented in
Sec.~\ref{sec2.2} and Appendix, but also the validity of two completely
independent numerical codes.

At this point, we must discuss one important aspect of the {\HF}
calculations in odd nuclei. In principle, for any given value of
$m_\lambda$, there may exist two solutions: one with prolate and
another one with oblate shape. Usually only the lowest one can be
converged; the other one, being excited, either does not converge or
falls down to the lowest one. In our calculations, in full agreement
with the standard Nilsson diagram~\cite{[RS80]}, we obtain converged
prolate (oblate) solutions for low-$m_\lambda$ (high-$m_\lambda$)
particle states, and {\it vise versa} for the hole states. We note
here that we did not constrain these solutions to axial symmetry;
nevertheless, stable triaxial solutions were never obtained.

\begin{figure}
\includegraphics[width=0.4\textwidth]{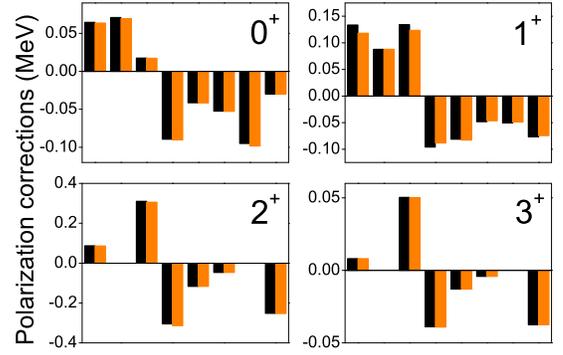}
\caption{(Color online) Polarization corrections of $|m_\lambda|=j_\lambda$
orbitals in $^{100}$Sn, determined by not including (left bars) and
including (right bars) the orbital-dependent terms in the RPA
matrices, see text. The order of orbitals is the same as shown
in Fig.~\protect\ref{fig:three}. Contributions coming from
four RPA channels $J^\pi=0^+$, 1$^+$, 2$^+$, and 3$^+$  are shown
separately (note very different scales).
}
\label{fig:one}
\end{figure}
Next, we tested the assumption, discussed in Sec.~\ref{sec2.2},
related to the smallness of terms $\rho^\lambda$ and $h^\lambda$ with
respect to the RPA expansion. In Fig.~\ref{fig:one}, we compare
polarization corrections determined by using the standard RPA
matrices with those containing the orbital-dependent terms [second
line in Eq.~(\ref{eq:ea})]. Since both sets of results are almost
identical, we conclude that in medium-heavy nuclei like $^{100}$Sn,
the orbital-dependent terms can be safely ignored. This significantly
simplifies the calculations, because a single common solution of the
RPA equation can then be used to determine polarization corrections
for all orbitals.

\begin{figure}
\includegraphics[width=0.4\textwidth]{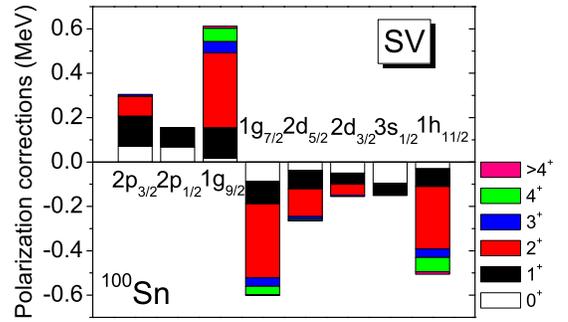}
\caption{(Color online) Contributions to
polarization corrections of $|m_\lambda|=j_\lambda$ orbitals in $^{100}$Sn,
coming from different $J^\pi$ RPA channels,
determined for the Skyrme EDF SV~\protect\cite{[Bei75]}.
The order of orbitals is the same as shown
in Fig.~\protect\ref{fig:three}. Contributions are ordered according
to the value of $J$, with the $0^+$ channels shown nearest the abscissa.
}
\label{fig:two}
\end{figure}
\begin{figure}
\includegraphics[width=0.4\textwidth]{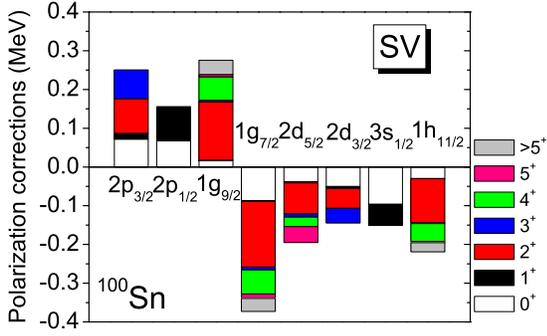}
\caption{(Color online) Same as in Fig.~\protect\ref{fig:two}
but for the $|m_\lambda|=\thalf$ orbitals.
}
\label{fig:two_two}
\end{figure}
In Fig.~\ref{fig:two}, we show
polarization corrections of the $|m_\lambda|=j_\lambda$ orbitals in $^{100}$Sn,
split into contributions from different $J^\pi$ RPA channels. First we note that the
geometric constraints in Eq.~(\ref{eq:86e}) limit the polarizations of
$j_\lambda$ orbitals to channels with $J\leq{2j_\lambda}$. As
expected, the largest contributions come from the coupling to the
quadrupole channel $2^+$, however, the monopole $0^+$ and dipole
$1^+$ channels also significantly contribute. For higher-$j_\lambda$
orbitals, channels $3^+$ and $4^+$ show some effect, whereas,
channels with $J>4$ can be safely neglected. For the $|m_\lambda|=\thalf$
orbitals shown in Fig.~\ref{fig:two_two}, the convergence is slightly slower,
but still all terms with $J>5$ contribute very little.

\subsection{RPA calculations in {\boldmath$^{100}$Sn\unboldmath} for the Skyrme EDF SLy5}
\label{sec3.2}

We now proceed to discuss the problem of SI energies in the EDF
calculations, presented in Sec.~\ref{sec2.4}. To this end, we
repeated the self-consistent calculations presented in
Sec.~\ref{sec3.1}, by employing the Skyrme EDF SLy5~\cite{[Cha98]}.
This is a standard Skyrme parametrization containing a strong density
dependent term, for which we can study the SI energies, as defined in
Eq.~(\ref{eq:88}). As before, the ``native'' time-odd terms of SLy5
were included and the center-of-mass correction to the kinetic energy
was neglected.

\begin{figure}
\includegraphics[width=0.4\textwidth]{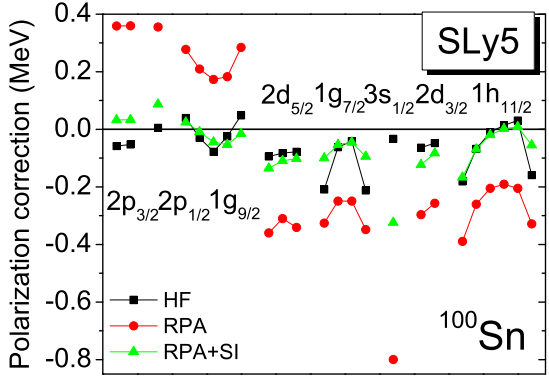}
\caption{(Color online) Same as in Fig.~\protect\ref{fig:three}, but for the
 Skyrme EDF SLy5~\protect\cite{[Cha98]}.
The RPA results correspond to the SIF terms in Eq.~(\protect\ref{eq:68}),
whereas RPA+SI denotes both SIF and SI contributions combined.
}
\label{fig:four}
\end{figure}
In Fig.~\ref{fig:four}, we show the RPA (SIF) contributions to
polarization corrections (\ref{eq:68}), and we compare the total
polarization corrections calculated by using Eq.~(\ref{eq:68}) with
the {\HF} results (\ref{eq:48}). The obtained agreement is very
good, although not as perfect as that obtained in Sec.~\ref{sec3.1}
for the Skyrme EDF SV. Moreover, the RPA results obtained for the SV and SLy5
functionals are significantly different from one another; the latter
ones being close to about $\pm$0.4\,MeV for holes and particles,
respectively. We also see that the SLy5 results are much less
$m_\lambda$-dependent.

\begin{figure}
\includegraphics[width=0.4\textwidth]{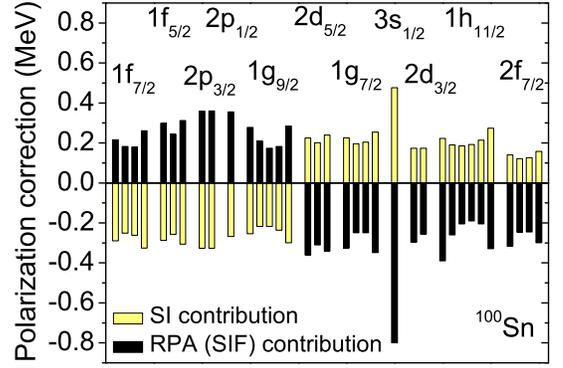}
\caption{(Color online) The SIF and SI contributions to the polarization corrections
of Eq.~(\protect\ref{eq:68}), calculated in $^{100}$Sn for the Skyrme EDF SLy5.
}
\label{fig:five}
\end{figure}
The most striking observation seen in Fig.~\ref{fig:four}, also explicitly
illustrated in Fig.~\ref{fig:five}, is a strong cancellation between
the SIF and SI contributions to the polarization corrections
(\ref{eq:68}). This cancellation makes the {\HF} polarization
corrections quite small, and gives the explanation to the long-standing
problem of significant differences between the magnitudes of the
{\HF} and RPA values~\cite{[Bor10]}. Indeed, it is the unphysical SI contribution
that renders the {\HF} polarization corrections so small, see
Ref.~\cite{[Zal08]} for a set of comprehensive calculations across
the mass chart.

\begin{figure}
\includegraphics[width=0.4\textwidth]{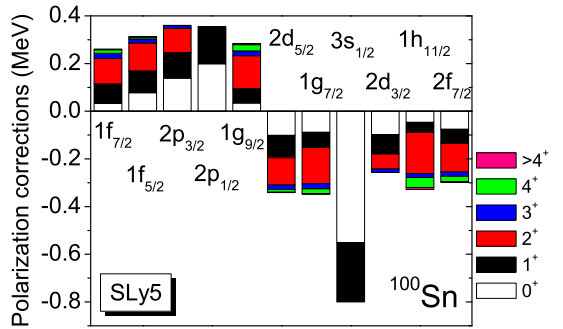}
\caption{(Color online) Same as in Fig.~\protect\ref{fig:two} but for
the contributions to the RPA SIF
polarization corrections of $|m_\lambda|=j_\lambda$ orbitals,
determined for the Skyrme EDF SLy5.
}
\label{fig:7a}
\end{figure}
\begin{figure}
\includegraphics[width=0.4\textwidth]{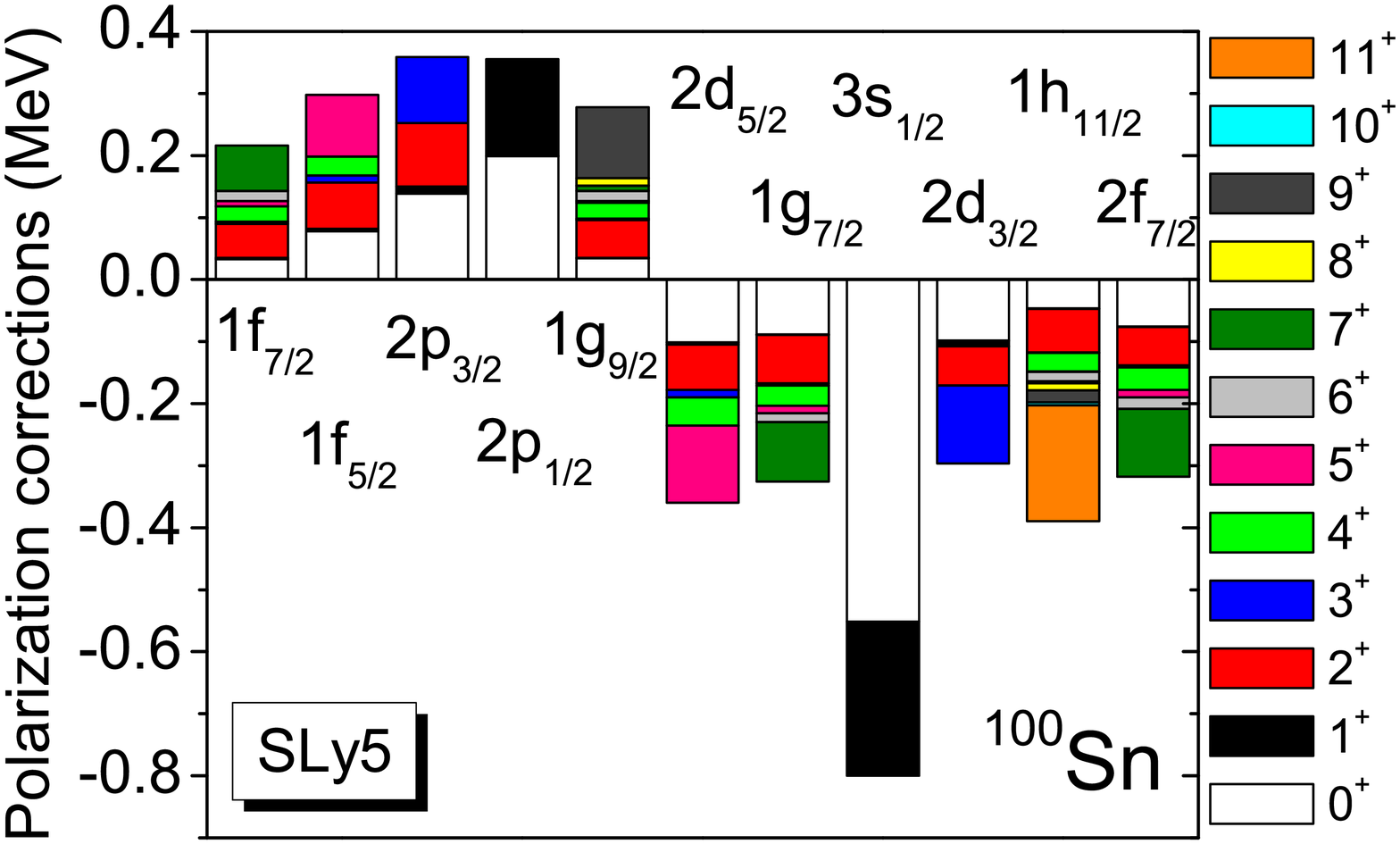}
\caption{(Color online) Same as in Fig.~\protect\ref{fig:7a}
but for the $|m_\lambda|=\thalf$ orbitals.
}
\label{fig:7b}
\end{figure}

\begin{figure*}
\includegraphics[width=0.9\textwidth]{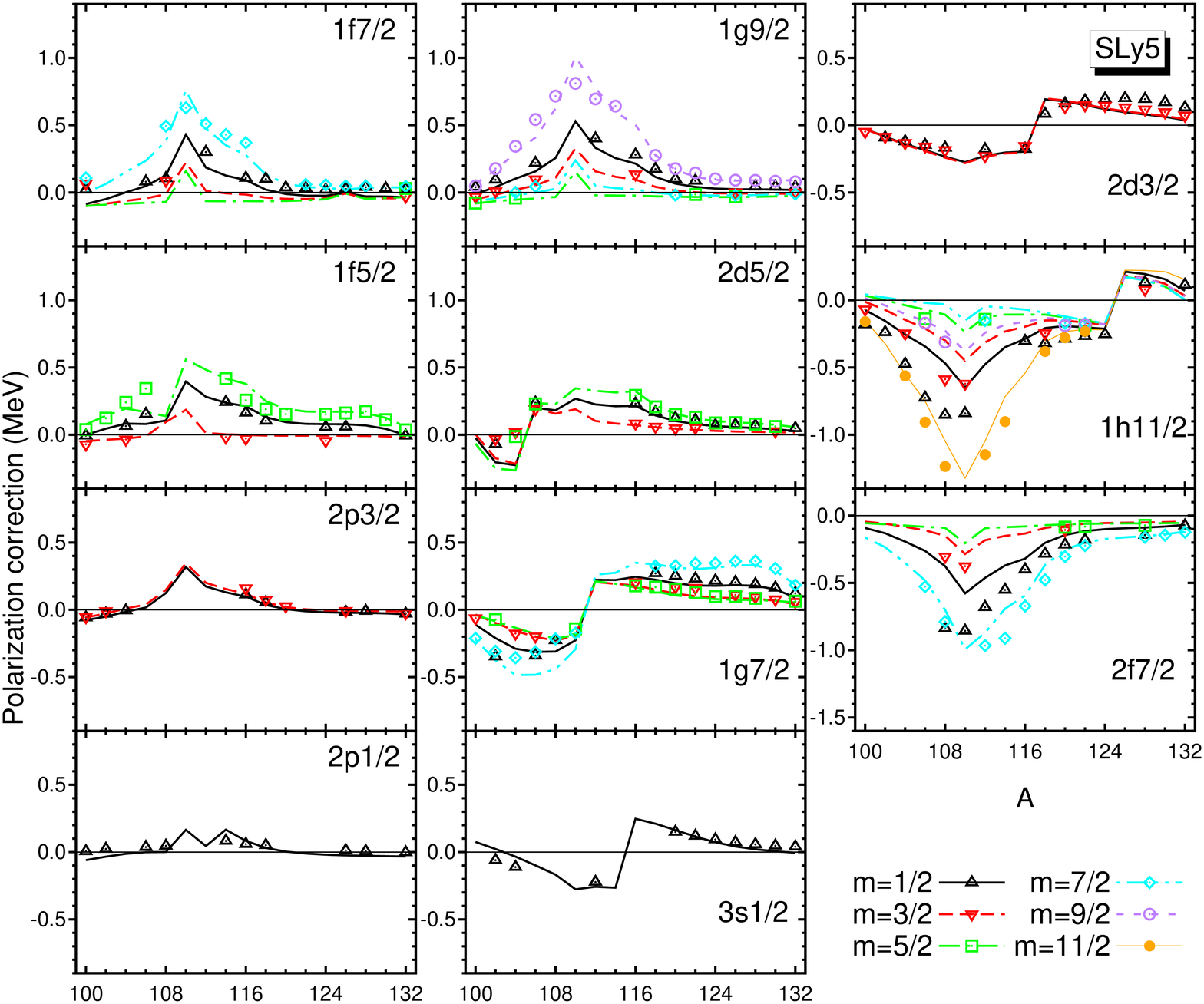}
\caption{(Color online) Comparison of the QRPA SIF+SI (symbols) and {\HFB} (lines)
polarization corrections to quasiparticle energies in tin isotopes.}
\label{fig:six}
\end{figure*}
We conclude this section by showing, in Figs.~\ref{fig:7a}
and~\ref{fig:7b}, the RPA SIF polarization corrections of the
$|m_\lambda|=j_\lambda$ and  $|m_\lambda|=\thalf$ orbitals,
respectively, split into contributions coming from different $J^\pi$ RPA
channels and calculated in $^{100}$Sn for the Skyrme EDF SLy5. These
can be compared with the analogous ones shown in Figs.~\ref{fig:two}
and~\ref{fig:two_two} for the Skyrme EDF SV. We first see that for
the $|m_\lambda|=j_\lambda$ orbitals, the convergence patterns
obtained for both EDFs are fairly similar. However, for the
$|m_\lambda|=\thalf$ orbitals, contributions coming from the
$J=2j_\lambda$ phonons turn out to be always quite large. For
example, results obtained for the h$_{11/2}$ orbital certainly require
taking into account the $J=11^+$ phonons.

\subsection{QRPA calculations in {\boldmath$^{100-132}$Sn\unboldmath} for the Skyrme EDF SLy5}
\label{sec3.3}

To present numerical results pertaining to the description of
polarization effects with pairing correlations included,
Sec.~\ref{sec2.3}, we performed {\HFB} and QRPA calculations for the
tin isotopes $^{100-132}$Sn. As in Sec.~\ref{sec3.2}, we used the
Skyrme EDF SLy5, whereas the pairing interaction was modelled by a
contact volume pairing force with the strength of
$V_0=200$\,MeV\,fm$^{-3}$ and active pairing space restricted to
states below 60\,MeV.

\begin{figure}
\includegraphics[width=0.4\textwidth]{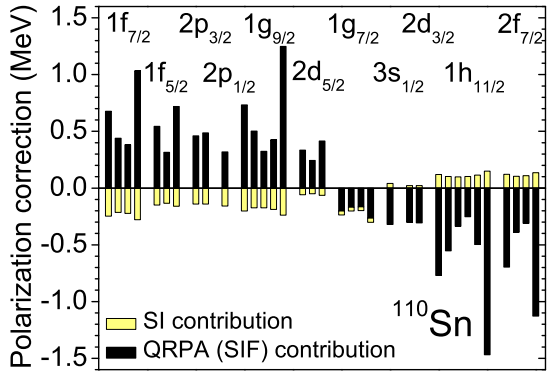}
\caption{(Color online) Same as in Fig.~\protect\ref{fig:five}
but for $^{110}$Sn.
}
\label{fig:9a}
\end{figure}

\begin{figure}
\includegraphics[width=0.4\textwidth]{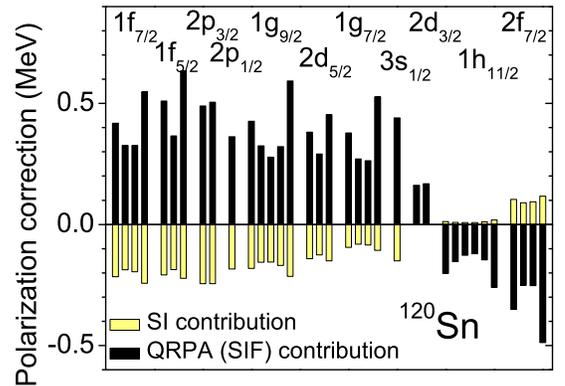}
\caption{(Color online) Same as in Fig.~\protect\ref{fig:five}
but for $^{120}$Sn.
}
\label{fig:9b}
\end{figure}

\begin{figure}
\includegraphics[width=0.4\textwidth]{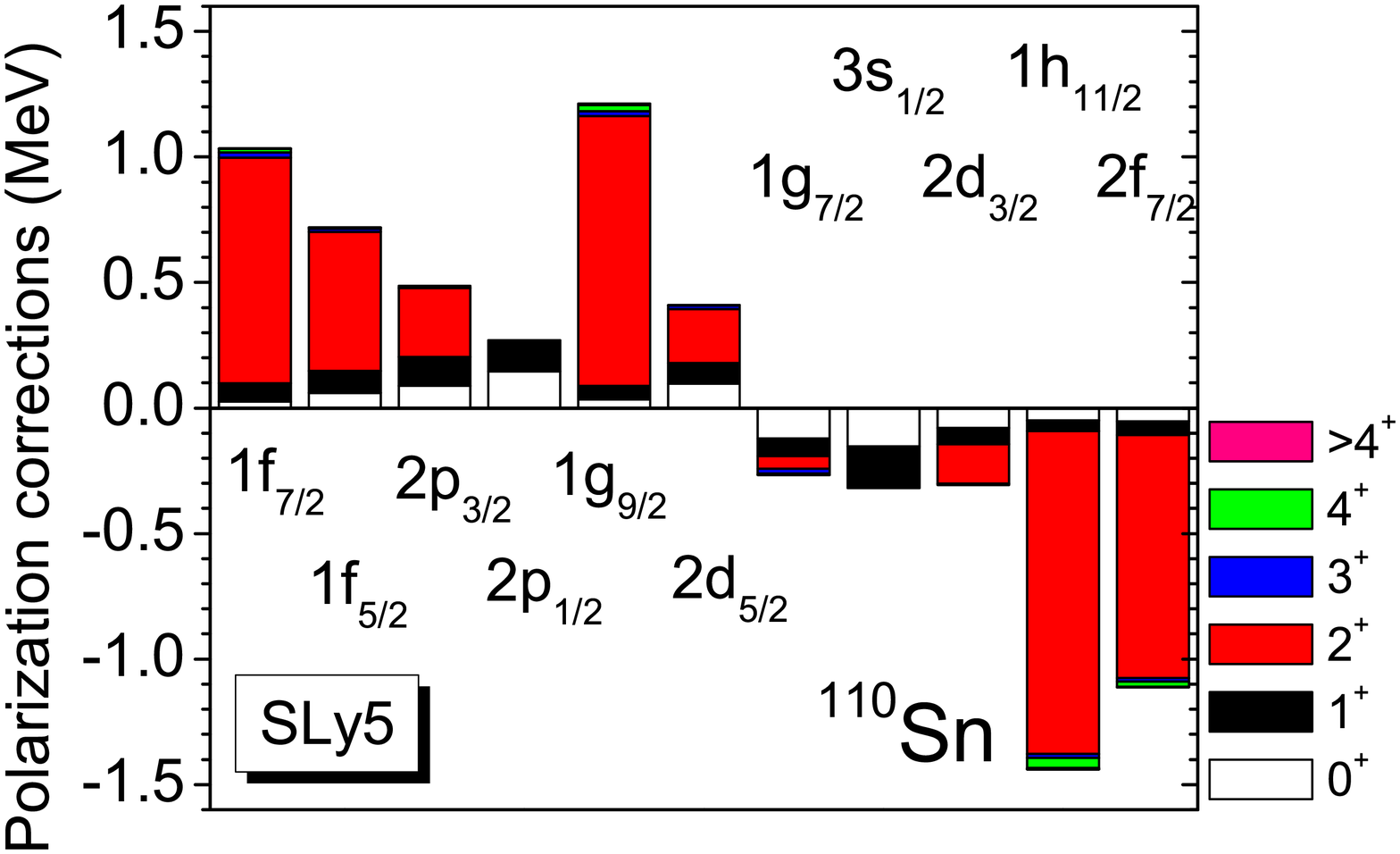}
\caption{(Color online) Same as in Fig.~\protect\ref{fig:7a}
but for $^{110}$Sn.
}
\label{fig:10a}
\end{figure}

\begin{figure}
\includegraphics[width=0.4\textwidth]{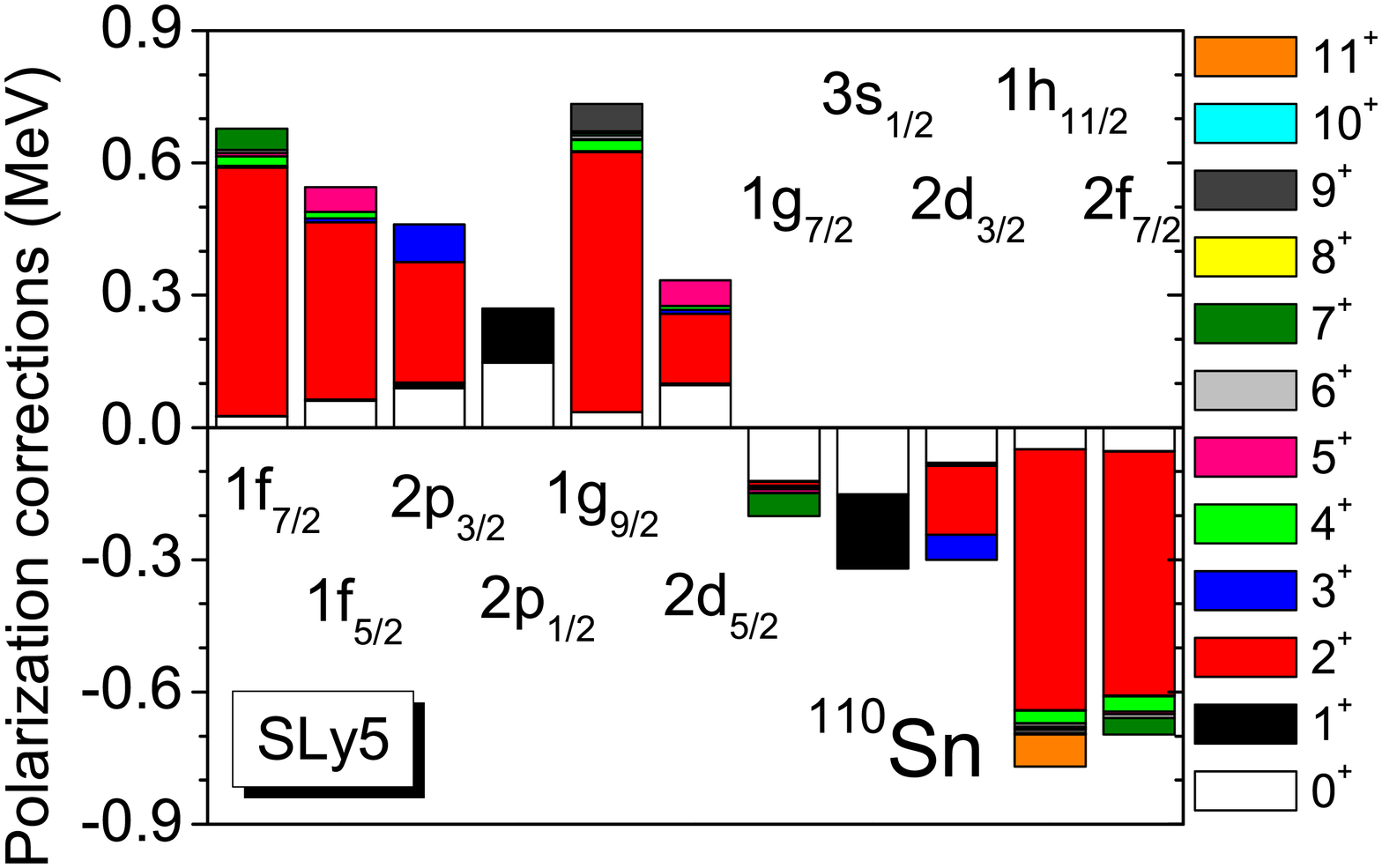}
\caption{(Color online) Same as in Fig.~\protect\ref{fig:7b}
but for $^{110}$Sn.
}
\label{fig:10b}
\end{figure}

\begin{figure}
\includegraphics[width=0.4\textwidth]{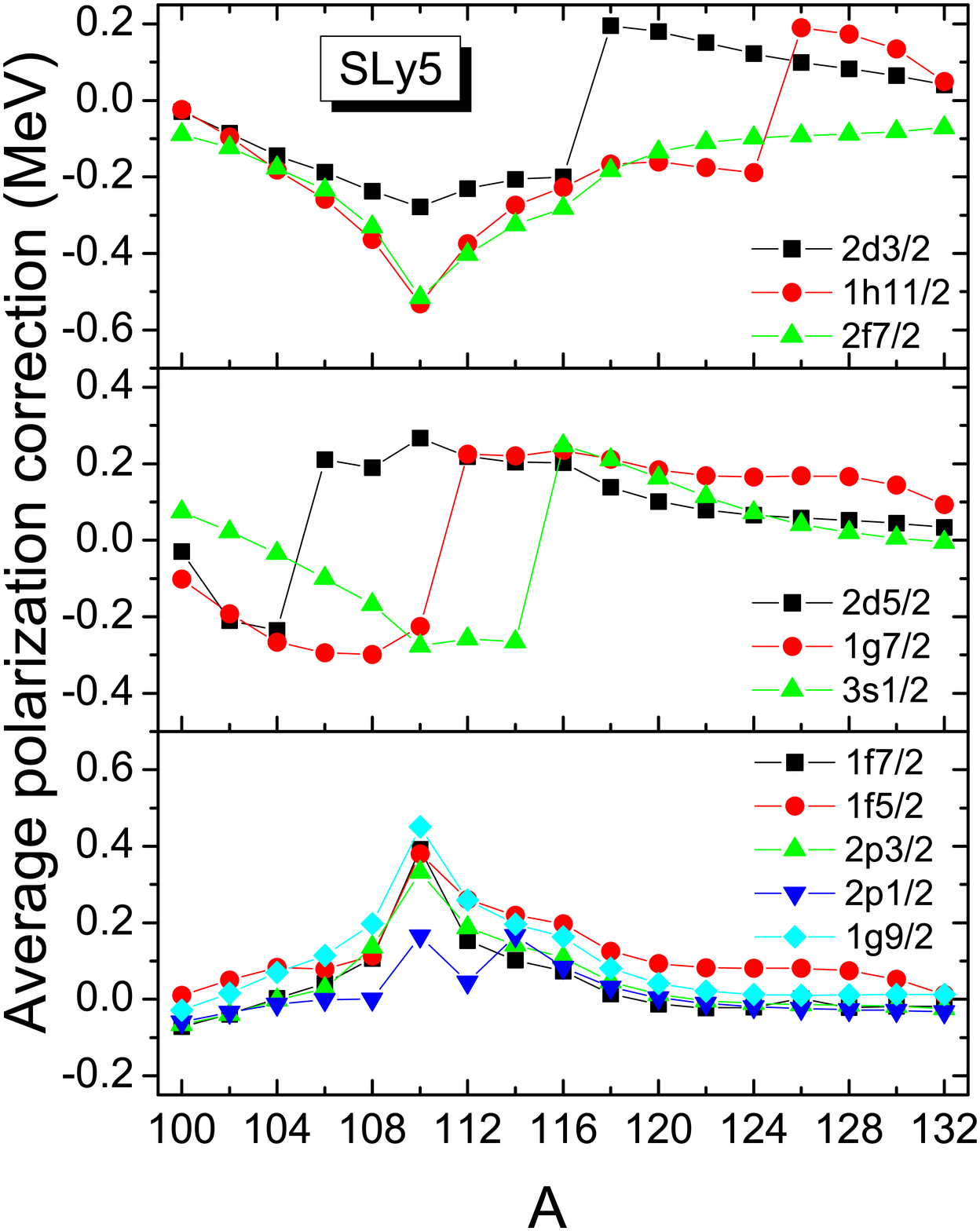}
\caption{(Color online) Average QRPA (SIF+SI) polarization corrections
$\delta E_{\text{SIF}}+E_{\text{SI}}$, Eqs.~(\protect\ref{eq:102}), in tin isotopes.
}
\label{fig:14}
\end{figure}

\begin{figure}
\includegraphics[width=0.4\textwidth]{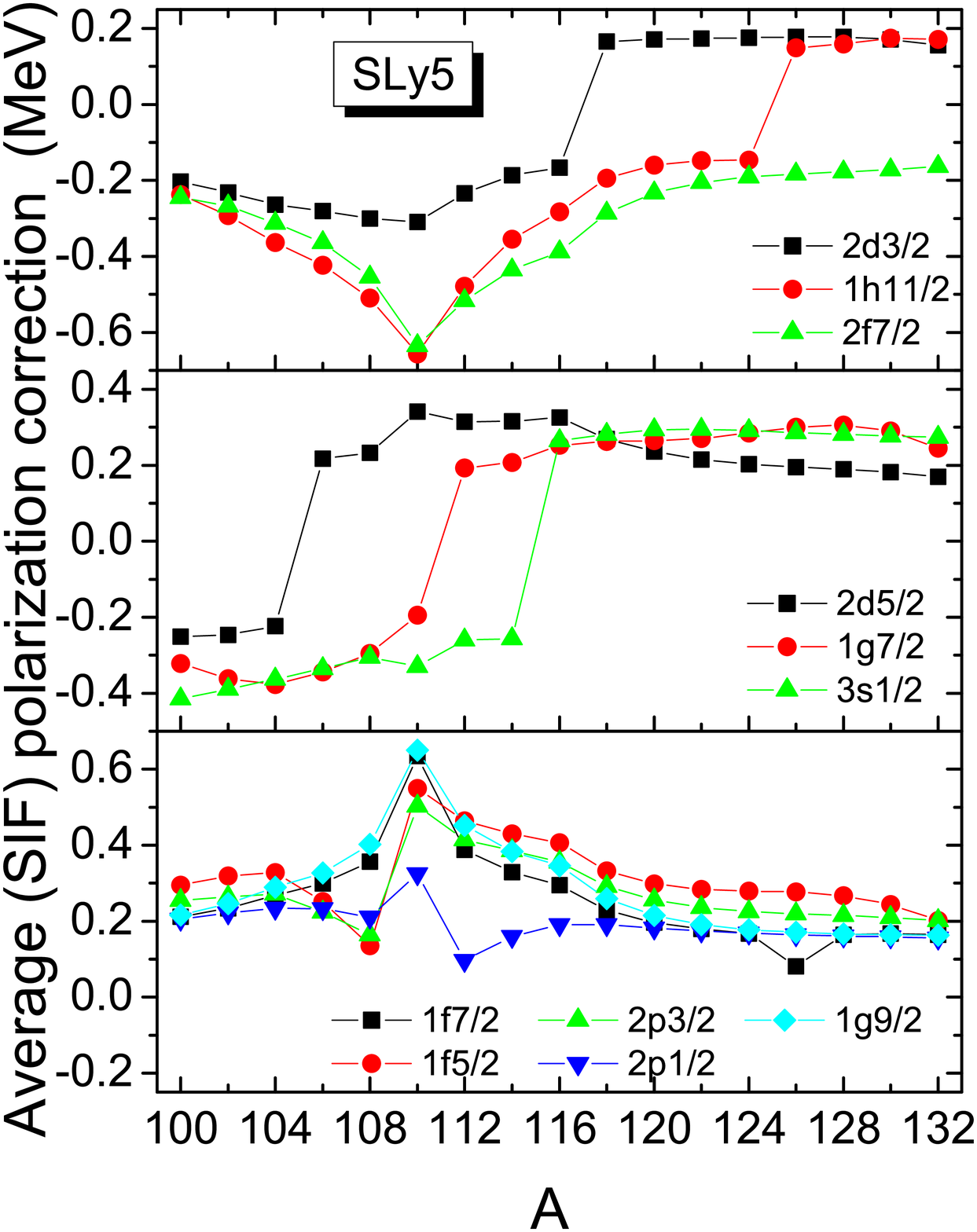}
\caption{(Color online) Same as in Fig.~\protect\ref{fig:14},
but for the QRPA SIF polarization corrections only.
}
\label{fig:15}
\end{figure}

\begin{figure}
\includegraphics[width=0.4\textwidth]{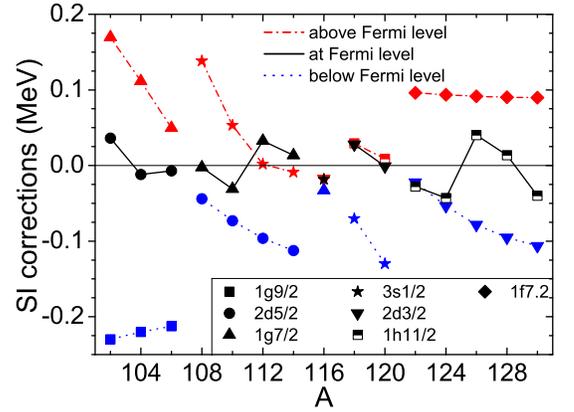}
\caption{(Color online) Average SI corrections $E_{\text{SI}}$
(\protect\ref{eq:102a}) plotted for quasiparticles closest
to the Fermi level in open-shell tin isotopes.
}
\label{fig:16}
\end{figure}

In Fig.~\ref{fig:six} we aim at testing Eq.~(\ref{eq:80}), where
$E^{A\pm1}$ and $E^A$ are self-consistent {\HFB} ground-state energies
of odd and even nuclei, respectively, $\lambda$ and $E_\Lambda$ are
the {\HFB} Fermi energy and quasiparticle energy of the blocked state,
and $\delta{}E$ is the QRPA (SIF+SI) polarization correction
(\ref{eq:86}). For this comparison, we must decide whether to use
the {\HFB} results obtained for the lighter $(A-1)$ or heavier $(A+1)$ odd
system.
Obviously, the former (latter) option must be used for predominantly
hole-type (particle-type) quasiparticles. For quasiparticles near the
Fermi level, however, there is a certain degree of ambiguity, which
we here arbitrarily resolve by checking whether the
single-particle energy $e_\Lambda$ corresponding to the blocked quasiparticle state
is below or above the Fermi level $\lambda$.
In practice, we determine $e_\Lambda$ by diagonalizing in the even nucleus the mean-field
Hamiltonian $h^A$, which is a part of the {\HFB} Hamiltonian (\ref{H-dens}).
In addition, to link results
presented in this section to those presented before for magic nuclei
without pairing, in Figs.~\ref{fig:six}--\ref{fig:16} we plot
results for hole states with flipped signs, that is,
\begin{subequations}
  \label{eq:80a}
\begin{eqnarray}
  \label{eq:80aa}
\!\!\!\!\!\! - \delta{}E &\!\!=\!\!& +(E^A - E^{A-1}) -(\lambda - E_\Lambda) ~~\mbox{for}~~ e_\Lambda<\lambda,  \\
  \label{eq:80ab}
\!\!\!\!\!\! + \delta{}E &\!\!=\!\!& -(E^A - E^{A+1}) -(\lambda + E_\Lambda) ~~\mbox{for}~~ e_\Lambda>\lambda,
\end{eqnarray}
\end{subequations}
[cf. Eq.~(\ref{eq:48})].

Within such a convention, in Fig.~\ref{fig:six} we show the QRPA SIF+SI
(symbols) and {\HFB} (lines) polarization corrections given by the
left-hand and right-hand sides of Eqs.~(\ref{eq:80a}), respectively.
We note that not all blocked quasiparticle states could be converged
in all studied nuclei, and thus in the figure there is quite a number
of missing {\HFB} points. Nevertheless, we conclude that the agreement
between the QRPA and {\HFB} results is satisfactory. By this we
establish the equivalence of the two methods in determining the
polarization corrections with pairing.

In Figs.~\ref{fig:9a} and~\ref{fig:9b}, we compare the QRPA SIF
(\ref{eq:86b}) and SI (\ref{eq:89a}) contributions to the polarization
corrections. Similarly as in the case without pairing, shown in Fig.~\ref{fig:five}, the
SIF and SI terms always have opposite signs, and thus the SI
partially cancels the SIF contribution. However, here the SI terms
are relatively smaller, and thus they to a lesser degree decrease the
SIF contributions, as compared to the results with no pairing.
It is fairly difficult to pin down specific reasons for the
qualitative differences between the SI energies obtained with and
without pairing correlations. It could be that the SI energies
related to density-dependence of the Skyrme interaction (\ref{eq:88})
and those related to differences between the pp and ph channels
(\ref{eq:90}), partially cancel out.

Convergence of the QRPA polarization corrections as a function of the
angular momentum $J$ of the QRPA phonons, shown in
Figs.~\ref{fig:10a} and~\ref{fig:10b}, is much faster than that without
pairing, cf.~Figs.~\ref{fig:7a} and~\ref{fig:7b}.
Here, the 2$^+$ channels clearly dominate. This can be interpreted as
the result of the quadrupole collectivity being increased by the pairing
correlations. In most cases, channels with $J>4$ can be safely neglected,
with the exception of the $J=2j_\Lambda$ channels that slightly
contribute to the corrections of the $m_\Lambda=\thalf$ quasiparticle
states.

All results presented up to now pertain to single-reference HF(B)
and (Q)RPA calculations, that is, only one single orbital, with a fixed
projection $m_\lambda$ or $m_\Lambda$, was occupied and was inducing
polarization effects. As discussed previously, this required symmetry
breaking in the HF(B) solutions and required coupling of (Q)RPA
phonons to odd particles in a symmetry-nonconserving way. However,
a symmetry-conserving (Q)RPA coupling~\cite{(Col10)} simply
amounts to averaging the results obtained for different values of
$m_\lambda$ or $m_\Lambda$, see Eqs.~(\ref{eq:102}). In
Figs.~\ref{fig:14}--\ref{fig:16}, we present results for the averages
obtained in this way.

Figures~\ref{fig:14} and \ref{fig:15} summarize our results obtained for
the QRPA polarization effects in tin isotopes. We see that the
polarization corrections strongly depend on $A$. This is mostly due
to the fact that for the Skyrme EDF SLy5, the quadrupole collectivity
in tin isotopes varies with $A$, and peaks near $A=110$ where the nuclei
are softest against the quadrupole deformation and the QRPA 2$^+$
phonons are lowest in energy and have the largest strength. At
$A=110$, values of polarization corrections reach up to 0.6\,MeV.

Note that when a given orbital crosses the Fermi level,
its plotted polarization correction jumps from negative to positive
values, which is the result of the plotting convention explained in
Eqs.~(\ref{eq:80a}). In fact, the QRPA polarization corrections to
quasiparticle energies are always negative. For the SIF
contributions, cf.~Eqs.~(\ref{eq:86b}) and (\ref{eq:86e}), this fact
is obvious, whereas for paired open-shell nuclei, smaller
opposite-sign SI contributions are unable to change this general
rule.

In Fig.~\ref{fig:16}, for selected quasiparticles in open-shell tin isotopes, we
show values of the average SI corrections $E_{\text{SI}}$
(\protect\ref{eq:102a}). Solid lines connect values obtained for
quasiparticles at the Fermi level, and dashed and dotted lines
pertain to those just below and above the Fermi level, respectively.
It is interesting to see that for the quasiparticle at the Fermi
level, the SI corrections become fairly small, not exceeding
50\,keV, whereas away from the Fermi level they can be of the order
of 200\,keV, see also Figs.~\ref{fig:9a} and~\ref{fig:9b}. If this
observation is confirmed or derived in a systematic way, we can hope
that near ground states of odd open-shell nuclei, the effects of SI
energies might be small. This is important, because the odd-even mass
staggering, where masses of odd nuclei enter, is routinely used to
gauge the intensity of pairing correlations.

\section{Conclusions}
\label{sec4}

In the present study, we investigated links between the mean-field
and polarization-correction approaches to masses of odd nuclei. The
former ones are rooted in the energy-density-functional methods and
strive to describe odd systems by blocking odd particle or
quasiparticle states. Energies of odd nuclei are then obtained by
minimization methods, in full analogy with those used in even nuclei,
and by employing the same energy density functionals. The latter ones
are based on the perturbative ``diagonal'' coupling between the odd
particle and vibrational phonons calculated in even systems.

Following the classic analyses presented in
Refs.~\cite{[Bro70],[Bla86],[Lip87],[Lip89]}, we derived links
between these two classes of approaches also in the case of
density-dependent interactions and/or paired systems. This allowed us
to show limitations of the polarization-correction methods as
compared to the full ``non-diagonal'' particle-vibration-coupling
that is rooted in the many-body perturbation theory.

We performed numerical analyses by using the deformed mean-field code
{\sc{hfodd}} (v2.52k)~\cite{[Sch12]}, which is able to solve
self-consistent equations in odd nuclei by breaking all symmetries.
The polarization corrections were independently calculated by using
the spherical code {\sc{hosphe}}~\cite{[Car13]}, which has the
capability to solve efficiently the (Q)RPA equations. The
comparison of results allowed us to identify the reason of
discrepancies between the masses of odd nuclei calculated with these
two approaches, which turns out to be the self-interaction energy,
polluting the mean-field energies of odd nuclei. Our derivations also
allowed us to explicitly define and calculate the self-interaction
energies, which then can be subtracted from the mean-field results
leading to self-interaction-free masses.

\begin{acknowledgments}

Fruitful and inspiring discussions with Gianluca Col\`o are
gratefully acknowledged. This work was supported in part by the THEXO
JRA within the EU-FP7-IA project ENSAR (No.\ 262010), Academy of
Finland and University of Jyv\"askyl\"a within the FIDIPRO programme,
Polish National Science Center under Contract No.\
2012/07/B/ST2/03907, and Bulgarian Science Fund under Contract No.\ NuPNET-SARFEN
DNS7RP01/0003. We acknowledge the CSC - IT Center for Science
Ltd, Finland, for the allocation of computational resources.

\end{acknowledgments}

\appendix

\section{Spherical symmetry}
\label{sec5}

In this Appendix, we specify the final equations obtained for the SI
energy (\ref{eq:89}) and polarization correction  (\ref{eq:86b}) to
the case of spherical symmetry, for which the numerical analyses of this
work were performed. First we note that the use of
spherical symmetry does not mean that in the odd system the spherical symmetry is
conserved.

Indeed, in the even system, the
quasiparticles move in the spherical field and thus their wave
functions are characterized by quantum numbers $\Lambda \equiv
n_\Lambda j_\Lambda m_\Lambda$ that comprise the principal quantum
number $n_\Lambda$, angular momentum $j_\Lambda$, and its projection
$m_\Lambda$. Quasiparticles having different projections $m_\Lambda$
are degenerate, and, therefore, any linear combination of them can be
used as the blocked orbital. In this work, we make a simplifying
assumption that the blocked quasiparticle corresponds to a specific single
value of the projection $m_\Lambda$. The general case could
have been treated equally easy, and at the end of the Appendix we
discuss the meaning of it.

An odd state, obtained by blocking a quasiparticle,
becomes necessarily deformed. In the
calculations performed with the deformed code {\sc{hfodd}}, this is
particularly well and explicitly visible, as the self-consistent
solutions obtained in odd systems always acquire small but non-zero
deformations. The aim of this Appendix is to show in which way the
deformation, and the dependence of final results on the values of
$m_\Lambda$, appears in the QRPA calculations that are performed in
the spherical basis and by using the spherical code {\sc{hosphe}}.

We begin by specifying expressions for the density matrix and pairing
tensor of the blocked quasiparticle (\ref{eq:kappdel3}) to the
case of the spherical basis $\alpha\equiv n_\alpha j_\alpha m_\alpha$,
\begin{subequations}
\label{eq:kappdel3a}
\begin{eqnarray}
  \rho_{n_\alpha j_\alpha m_\alpha,n_\beta j_\beta m_\beta}^\Lambda  &=&-V_{n_\alpha j_\alpha m_\alpha,n_\Lambda j_\Lambda m_\Lambda}^{*}V_{n_\beta j_\beta m_\beta,n_\Lambda j_\Lambda m_\Lambda} \nonumber \\&&+U_{n_\alpha j_\alpha m_\alpha,n_\Lambda j_\Lambda m_\Lambda}U_{n_\beta j_\beta m_\beta,n_\Lambda j_\Lambda m_\Lambda}^{*} , \nonumber \\ \\
\kappa_{n_\alpha j_\alpha m_\alpha,n_\beta j_\beta m_\beta}^\Lambda  &=&-V_{n_\alpha j_\alpha m_\alpha,n_\Lambda j_\Lambda m_\Lambda}^{*}U_{n_\beta j_\beta m_\beta,n_\Lambda j_\Lambda m_\Lambda} \nonumber \\&&+U_{n_\alpha j_\alpha m_\alpha,n_\Lambda j_\Lambda m_\Lambda}V_{n_\beta j_\beta m_\beta,n_\Lambda j_\Lambda m_\Lambda}^{*} , \nonumber \\
\end{eqnarray}
\end{subequations}
where the spherically-symmetric quasiparticle wave functions read,
\begin{subequations}
\label{spherical}
\begin{eqnarray}
\label{spherical3} \!\!\!\! \!\!\!\!  \!\!\!\! \!\!\!\!
  U_{n_\alpha j_\alpha m_\alpha,n_L j_L m_L}
  &\!\!=\!\!&  \delta_{j_\alpha j_L}\delta_{m_\alpha  m_L} U^{j_L}_{n_\alpha n_L} , \\
\label{spherical4} \!\!\!\! \!\!\!\! \!\!\!\! \!\!\!\!
  V_{n_\alpha j_\alpha m_\alpha,n_L j_L m_L}
  &\!\!=\!\!&   (-1)^{j_\alpha-m_\alpha}
       \delta_{j_\alpha j_L}\delta_{m_\alpha,-m_L} V^{j_L}_{n_\alpha n_L} .
\end{eqnarray}
\end{subequations}
and $U^{j_L}_{n_\alpha n_L}$ and $V^{j_L}_{n_\alpha n_L}$ are solutions of
the {\HFB} equation, obtained for the quasiparticle state with quantum
numbers $L\equiv n_L j_L m_L$ in the given $j_L$ block.

Similarly as for the angular-momentum-projected deformed
states~\cite{[RS80]}, we can write the deformed density matrix
$\rho^\Lambda$ and pairing tensor $\kappa^\Lambda$ as sums of those
projected on good angular momentum $J$ and its projections on
the laboratory axis $M$ and on the intrinsic axis $K$,
$\rho^{\Lambda,JMK}$ and $\kappa^{\Lambda,JMK}$, that is,
\begin{subequations}
\label{eq:98}
\begin{eqnarray}
\label{eq:98e}
  \rho_{n_\alpha j_\alpha m_\alpha,n_\beta j_\beta m_\beta}^\Lambda
&=& \sum_{JK}
  \rho_{n_\alpha j_\alpha m_\alpha,n_\beta j_\beta m_\beta}^{\Lambda,JKK}
, \\
\label{eq:98f}
\kappa_{n_\alpha j_\alpha m_\alpha,n_\beta j_\beta m_\beta}^\Lambda
&=& \sum_{JK}
  \kappa_{n_\alpha j_\alpha m_\alpha,n_\beta j_\beta m_\beta}^{\Lambda,JKK}
,
\end{eqnarray}
\end{subequations}
where only the $M=K$ terms appear in the expansion~\cite{[RS80]}.
By using the Wigner-Eckart theorem~\cite{[Var88]}, one can always express
laboratory spherical-tensor matrices,
$\rho^{\Lambda,JMK}$ and $\kappa^{\Lambda,JMK}$,
corresponding to quantum numbers $JM$, as,
\begin{subequations}
\begin{eqnarray}
\label{eq:98c}
  \rho_{n_\alpha j_\alpha m_\alpha,n_\beta j_\beta m_\beta}^{\Lambda,JMK}
  &=&
\frac{1}{\sqrt{2j_\alpha+1}}C^{j_\alpha m_\alpha}_{j_\beta m_\beta JM}
\nonumber \\ &&\times
\langle{n_\alpha j_\alpha}||\rho^{\Lambda,JK}||{n_\beta j_\beta}\rangle
, \\
\label{eq:98d}
   \kappa_{n_\alpha j_\alpha m_\alpha,n_\beta j_\beta m_\beta}^{\Lambda,JMK}
  &=&
\frac{1}{\sqrt{2j_\alpha+1}}(-1)^{j_\beta-m_\beta}
C^{j_\alpha m _\alpha}_{j_\beta,-m _\beta JM}
\nonumber \\ &&\times
\langle{n_\alpha j_\alpha}||\kappa^{\Lambda,JK}||{n_\beta j_\beta}\rangle
,
\end{eqnarray}
\end{subequations}
where the reduced matrix elements can be calculated as,
\begin{subequations}
\label{eq:94}
\begin{eqnarray}
\label{eq:94a}
\langle{n_\alpha j_\alpha}||\rho^{\Lambda,JK}||{n_\beta j_\beta}\rangle
  &=& \sum_{m_\alpha m_\beta}\frac{2J+1}{\sqrt{2j_\alpha+1}}
C^{j_\alpha m_\alpha}_{j_\beta m_\beta JK}
\nonumber \\ &&\times
\rho_{n_\alpha j_\alpha m_\alpha,n_\beta j_\beta m_\beta}^\Lambda, \\
\label{eq:94b}
\langle{n_\alpha j_\alpha}||\kappa^{\Lambda,JK}||{n_\beta j_\beta}\rangle
  &=& \sum_{m_\alpha m_\beta}\frac{2J+1}{\sqrt{2j_\alpha+1}}
C^{j_\alpha m_\alpha}_{j_\beta m_\beta JK}
\nonumber \\ &&\hspace*{-2cm}\times
(-1)^{j_\beta+m_\beta} \kappa_{n_\alpha j_\alpha m_\alpha,n_\beta j_\beta,-m_\beta}^\Lambda.
\end{eqnarray}
\end{subequations}
Validity of expansions (\ref{eq:98}) can now be explicitly verified
by using summation properties of the Clebsh-Gordan coefficients~\cite{[Var88]}.

At this point, we can use the fact that the spherical-basis properties of
mean fields are exactly the same as those of densities, that is,
Eqs.~(\ref{eq:98})--(\ref{eq:94}) hold equally well for $\rho^\Lambda$ and
$\kappa^\Lambda$ replaced by $\Gamma^\Lambda$ and $\Delta^\Lambda$,
respectively. Then, by summing the Clebsh-Gordan coefficients again,
traces in Eq.~(\ref{eq:89}) can be explicitly evaluated, which gives,
\begin{eqnarray}
  \label{eq:101e} \hspace*{-1cm}
E^\Lambda_{\text{SI}}
 &=&  \tfrac{1}{2}\sum_{n_\alpha j_\alpha n_\beta j_\beta JK}
\frac{1}{2J+1} \nonumber \\
&&\hspace*{-1cm}\times\Big(
\langle{n_\alpha j_\alpha}||\Gamma^{\Lambda,JK}||{n_\beta j_\beta}\rangle
\langle{n_\alpha j_\alpha}||\rho^{\Lambda,J,-K}||{n_\beta j_\beta}\rangle^* \nonumber \\
 &&\hspace*{-1cm} ~ +
\langle{n_\alpha j_\alpha}||\Delta^{\Lambda,JK}||{n_\beta j_\beta}\rangle
\langle{n_\alpha j_\alpha}||\kappa^{\Lambda,JK}||{n_\beta j_\beta}\rangle^*\Big) .
\end{eqnarray}

Similarly, we can evaluate the QRPA SIF polarization correction of
Eq.~(\ref{eq:86b}).
Since the spherical QRPA amplitudes $X$ and $Y$
can be labeled with the good quantum numbers $JM$, we have
\begin{widetext}
\begin{eqnarray}
  \label{eq:86c}
\delta E^{\Lambda}_{\text{SIF}}
   &=&  -\tfrac{1}{4}\sum_{JM}
    \sum_{\omega_J>0}\frac{\left\vert
   \sum_{LL'}\left(W^{\Lambda*}_{LL'}X_{LL'}^{\omega,JM}
                   +W^{\Lambda }_{LL'}Y_{LL'}^{\omega,JM}
              \right)\right\vert ^{2}}{\hbar\omega_J} .
\end{eqnarray}
From the Wigner-Eckart theorem, amplitudes $X^{\omega,JM}$ and $Y^{\omega,JM}$ read
\begin{eqnarray}
\label{eq:109b}
X_{n_L j_L m_L,n_{L'} j_{L'} m_{L'}}^{\omega,JM}
\!\!&\!\!=\!\!&\!\!
\frac{1}{\sqrt{2j_L+1}}(-1)^{j_{L'}-m_{L'}}
C^{j_L m _L}_{j_{L'},-m _{L'} JM}
\langle{n_L j_L}||X^{\omega,J}||{n_{L'} j_{L'}}\rangle
,
\\
\label{eq:109c}
Y_{n_L j_L m_L,n_{L'} j_{L'} m_{L'}}^{\omega,JM}
\!\!&\!\!=\!\!&\!\!
\frac{1}{\sqrt{2j_L+1}}(-1)^{j_L-m_L}
C^{j_L,-m _L}_{j_{L'} m _{L'} JM}
\langle{n_L j_L}||Y^{\omega,J}||{n_{L'} j_{L'}}\rangle
.
\end{eqnarray}
By using the fact that the spherical-basis properties of quasiparticle
matrix $W^\Lambda$ are the same as those of $\kappa^\Lambda$, see
Eqs.~(\ref{eq:98f}),  (\ref{eq:98d}), and (\ref{eq:94b}), we can derive that
\begin{eqnarray}
  \label{eq:86a}
\delta E^{\Lambda}_{\text{SIF}}
  &=&-\tfrac{1}{4}
        \sum_{JK}\frac{1}{(2J+1)^2}\sum_{\omega_{J}>0}\frac{1}{\hbar\omega_{J}}\Bigg\vert
        \sum_{n_L j_L n_{L'} j_{L'}}\Big(
\langle{n_L j_L}||W^{\Lambda,JK}||{n_{L'} j_{L'}}\rangle^*
\langle{n_L j_L}||X^{\omega,J} ||{n_{L'} j_{L'}}\rangle
\nonumber \\  && \hspace*{4cm}
 - (-1)^{J+K}
\langle{n_L j_L}||W^{\Lambda,J,-K} ||{n_{L'} j_{L'}}\rangle
\langle{n_L j_L}||Y^{\omega,J}||{n_{L'} j_{L'}}\rangle
\Big)
              \Bigg\vert^{2} .
\end{eqnarray}
\end{widetext}

Finally, we note that Eqs.~(\ref{eq:98})--(\ref{eq:86a}) hold for
an arbitrary blocked quasiparticle. However, when the
reduced matrix elements (\ref{eq:94}) are evaluated for
a specific quasiparticle (\ref{spherical}) that has
a fixed value of projection $m_\Lambda$, we obtain,
\begin{subequations}
\label{eq:96}
\begin{eqnarray}
\label{eq:96a}
\langle{n_\alpha j_\alpha}||\rho^{JK}||{n_\beta j_\beta}\rangle
  &=&\frac{2J+1}{\sqrt{2j_\Lambda+1}}C^{j_\Lambda m_\Lambda}_{j_\Lambda m_\Lambda J0}
\delta_{j_\alpha,j_\Lambda}\delta_{j_\beta ,j_\Lambda}\delta_{K0} \nonumber \\
  && \hspace*{-3cm} \times
          \left(-(-1)^J V^{j_\Lambda*}_{n_\alpha n_\Lambda}V^{j_\Lambda }_{n_\beta n_\Lambda}
                      + U^{j_\Lambda }_{n_\alpha n_\Lambda}U^{j_\Lambda*}_{n_\beta n_\Lambda}\right), \\
\label{eq:96b}
\langle{n_\alpha j_\alpha}||\kappa^{JK}||{n_\beta j_\beta}\rangle
  &=&\frac{2J+1}{\sqrt{2j_\Lambda+1}}C^{j_\Lambda m_\Lambda}_{j_\Lambda m_\Lambda J0}
\delta_{j_\alpha,j_\Lambda}\delta_{j_\beta ,j_\Lambda}\delta_{K0} \nonumber \\
 && \hspace*{-3cm} \times
          \left( (-1)^J V^{j_\Lambda*}_{n_\alpha n_\Lambda}U^{j_\Lambda }_{n_\beta n_\Lambda}
                      + U^{j_\Lambda }_{n_\alpha n_\Lambda}V^{j_\Lambda*}_{n_\beta n_\Lambda}\right).
\end{eqnarray}
\end{subequations}
In this case, in Eqs.~(\ref{eq:101e}) and  (\ref{eq:86a}) only the
$K=0$ terms contribute to the SI energy and  SIF polarization correction,
respectively.

In any channel $J$, the results depend on $m_\Lambda$ only
through the Clebsh-Gordan coefficient
$C^{j_\Lambda m_\Lambda}_{j_\Lambda m_\Lambda J0}$.
It is, therefore, advantageous to define triple reduced
matrix elements that do not depend on $m_\Lambda$,
\begin{subequations}
\label{eq:107}
\begin{eqnarray}
\label{eq:107a}
\langle{\alpha j_\alpha}||\rho^{\Lambda,JK}||{\beta j_\beta}\rangle
  &\!\!=\!\!&F(m_\Lambda,J)
\delta_{K0}
\langle{\alpha j_\alpha}|||\rho^{\Lambda,J}|||{\beta j_\beta}\rangle, \nonumber \\  \\
\label{eq:107b}
\langle{\alpha j_\alpha}||\kappa^{\Lambda,JK}||{\beta j_\beta}\rangle
  &\!\!=\!\!&F(m_\Lambda,J)
\delta_{K0}
\langle{\alpha j_\alpha}|||\kappa^{\Lambda,J}|||{\beta j_\beta}\rangle ,  \nonumber \\
\end{eqnarray}
\end{subequations}
for
\begin{equation}
\label{eq:107c}
F(m_\Lambda,J)=\sqrt{2J+1}C^{j_\Lambda m_\Lambda}_{j_\Lambda m_\Lambda J0}.
\end{equation}
Then, by using the triple reduced
matrix elements, calculation of mean fields can be performed only once,
and the results valid for arbitrary values of $m_\Lambda$ can be
reconstructed as,
\begin{subequations}
\label{eq:117x}
\begin{eqnarray}
\label{eq:117a}
\langle{\alpha j_\alpha}||\Gamma^{\Lambda,JK}||{\beta j_\beta}\rangle
  &\!\!=\!\!&F(m_\Lambda,J)
\delta_{K0}
\langle{\alpha j_\alpha}|||\Gamma^{\Lambda,J}|||{\beta j_\beta}\rangle, \nonumber \\ \\
\label{eq:117b}
\langle{\alpha j_\alpha}||\Delta^{\Lambda,JK}||{\beta j_\beta}\rangle
  &\!\!=\!\!&F(m_\Lambda,J)
\delta_{K0}
\langle{\alpha j_\alpha}|||\Delta^{\Lambda,J}|||{\beta j_\beta}\rangle, \nonumber \\ \\
\label{eq:117c}
\langle{\alpha j_\alpha}||W^{\Lambda,JK}||{\beta j_\beta}\rangle
  &\!\!=\!\!&F(m_\Lambda,J)
\delta_{K0}
\langle{\alpha j_\alpha}|||W^{\Lambda,J}|||{\beta j_\beta}\rangle. \nonumber \\
\end{eqnarray}
\end{subequations}

In terms of the triple reduced
matrix elements, for fixed-$m_\Lambda$ quasiparticles, results
(\ref{eq:101e}) and (\ref{eq:86a}) can be expressed through
contributions coming from different channels,
\begin{widetext}
\begin{eqnarray}
  \label{eq:101f} \hspace*{-1cm}
E^\Lambda_{\text{SI},J}
&=&\tfrac{1}{2}\sum_{n_\alpha n_\beta}\Big(
\langle{n_\alpha j_\Lambda}|||\Gamma^{\Lambda,J}|||{n_\beta j_\Lambda}\rangle
\langle{n_\alpha j_\Lambda}|||\rho^{\Lambda,J}|||{n_\beta j_\Lambda}\rangle^*
  +
\langle{n_\alpha j_\Lambda}|||\Delta^{\Lambda,J}|||{n_\beta j_\Lambda}\rangle
\langle{n_\alpha j_\Lambda}|||\kappa^{\Lambda,J}|||{n_\beta j_\Lambda}\rangle^*\Big) , \\
  \label{eq:86e}
\delta E^{\Lambda}_{\text{SIF},J}
  &=&-\frac{1}{4(2J+1)}
\sum_{\omega_{J}>0}\frac{1}{\hbar\omega_{J}}\Bigg\vert
        \sum_{n_L j_L n_{L'} j_{L'}}\Big(
\langle{n_L j_L}|||W^{\Lambda,J}|||{n_{L'} j_{L'}}\rangle^*
\langle{n_L j_L}||X^{\omega,J} ||{n_{L'} j_{L'}}\rangle
\nonumber \\  && \hspace*{3cm}
 - (-1)^{J}
\langle{n_L j_L}|||W^{\Lambda,J} |||{n_{L'} j_{L'}}\rangle
\langle{n_L j_L}||Y^{\omega,J}||{n_{L'} j_{L'}}\rangle
\Big)
              \Bigg\vert^{2} ,
\end{eqnarray}
\end{widetext}
whereupon they read,

\begin{subequations}
  \label{eq:101faa}
\begin{eqnarray}
  \label{eq:101fa}
E^\Lambda_{\text{SI}}
 &=&  \sum_{J}\left(C^{j_\Lambda m_\Lambda}_{j_\Lambda m_\Lambda J0}\right)^2
E^\Lambda_{\text{SI},J} , \\
  \label{eq:86ea}
\delta E^{\Lambda}_{\text{SIF}}
  &=&   \sum_{J}\left(C^{j_\Lambda m_\Lambda}_{j_\Lambda m_\Lambda J0}\right)^2
\delta E^{\Lambda}_{\text{SIF},J} .
\end{eqnarray}
\end{subequations}

We note that factors
$\delta_{j_\alpha,j_\Lambda}\delta_{j_\beta,j_\Lambda}$, which are
present in Eqs.~(\ref{eq:96}), allowed for reducing
Eq.~(\ref{eq:101f}) to terms with $j_\alpha=j_\beta=j_\Lambda$ only.
However, fields (\ref{eq:117x}) are not restricted to $j_L=j_{L'}=j_\Lambda$
and thus the QRPA SIF corrections (\ref{eq:86e}) must be summed up
over $j_L$ and $j_{L'}$.

From expressions (\ref{eq:101faa}) we see that the
polarization corrections calculated for orbitals with given values of
$m_\Lambda$ are obtained by folding the $J$-dependent (but
$m_\Lambda$-independent) contributions $E^\Lambda_{\text{SI},J}$ and
$\delta E^{\Lambda}_{\text{SIF},J}$ with simple Clebsh-Gordan
coefficients. Values of these coefficients thus dictate how strongly
a given channel $J$ contributes. Moreover, owing to the summation
properties of the Clebsh-Gordan coefficients, the average contributions read,

\begin{subequations}
  \label{eq:102}
\begin{eqnarray}
  \label{eq:102a}
E_{\text{SI}}\equiv
\frac{1}{2j_\Lambda+1}\sum_{m_\Lambda}E^\Lambda_{\text{SI}}
 &=&  \sum_{J} \frac{1}{2J+1}
E^\Lambda_{\text{SI},J} , \nonumber \\ \\
  \label{eq:102b}
\delta E_{\text{SIF}}\equiv
\frac{1}{2j_\Lambda+1}\sum_{m_\Lambda}\delta E^\Lambda_{\text{SIF}}
 &=&  \sum_{J} \frac{1}{2J+1}
\delta E^\Lambda_{\text{SIF},J} .  \nonumber \\
\end{eqnarray}
\end{subequations}

As shown in this Appendix, by blocking quasiparticles that have fixed
values of projections $m_\Lambda$, one obtains only the $K=0$ terms
in densities and fields, that is, deformations of odd systems are
axial. It is also clear that by blocking quasiparticles with mixed
values of $m_\Lambda$, one would have obtained non-zero reduced
matrix elements for non-zero values of $K$, and thus in odd systems,
non-axial deformations would have appeared. Numerical results
presented in this study indicate, however, that axial solutions
have systematically lower energies.


\end{document}